# Decoding Structure-Spectrum Relationships with Physically Organized Latent Spaces

*Zhu Liang*,[1] *Matthew R. Carbone*,[2] *Wei Chen*,[1] *Fanchen Meng*,[1] *Eli Stavitski*,[3] *Deyu Lu*,[1,*] *Mark S. Hybertsen*,[1,*] and *Xiaohui Qu*[1,*]

[1]Center for Functional Nanomaterials, Brookhaven National Laboratory, Upton, New York 11973, USA

[2]Computational Science Initiative, Brookhaven National Laboratory, Upton, New York 11973, USA

[3]National Synchrotron Light Source II, Brookhaven National Laboratory, Upton, New York 11973, USA

* dlu@bnl.gov, mhyberts@bnl.gov, xiaqu@bnl.gov

## ABSTRACT

A new semi-supervised machine learning method for the discovery of structure-spectrum relationships is developed and demonstrated using the specific example of interpreting X-ray absorption near-edge structure (XANES) spectra. This method constructs a one-to-one mapping between individual structure descriptors and spectral trends. Specifically, an adversarial autoencoder is augmented with a novel "rank constraint" (RankAAE). The RankAAE methodology produces a continuous and interpretable latent space, where each dimension can track an individual structure descriptor. As a part of this process, the model provides a robust and quantitative measure of the structure-spectrum relationship by decoupling intertwined spectral contributions from multiple structural characteristics. This makes it ideal for spectral interpretation and the discovery of new descriptors. The capability of this procedure is showcased by considering five local structure descriptors and a database of over fifty thousand simulated XANES spectra across eight first-row transition metal oxide families. The resulting structure-spectrum relationships not only reproduce known trends in the literature, but also reveal unintuitive ones that are visually indiscernible in large data sets. The results suggest that the RankAAE methodology has great potential to assist researchers to interpret complex scientific data, test physical hypotheses, and reveal new patterns that extend scientific insight.

## I. INTRODUCTION

Structure-property relationships in materials encode fundamental physical knowledge and enable a constructive design process to meet functional goals and guide materials discovery.[1–3] Considering materials beyond the simplest of molecules or crystals, the full description of the atomic-scale structure generally involves many separate quantities. Similarly, a collection of properties can also be a complex data set. Consequently, the discovery and representation of structure-property relationships pose significant challenges in the raw form of a "many-to-many" map. Traditionally, expert intuition has been used to identify a few simple structure descriptors that can be related to specific trends in properties through physical arguments and experimental probes.[4] The emergence of high-throughput experiments and expanding databases of computed materials properties invites the application of new, data-driven approaches to discover deeper, more comprehensive structure-property relationships.[1–3] Significant progress has been made in applying machine learning methods to physical datasets in recent years.[1–3,5–7] However, the development of interpretable models,[8–12] highly desirable for use by physical science domain experts, remains an ongoing challenge in the field.

Mapping the trends in a complex dataset, such as a collection of physical spectra measured for a set of materials, to a reduced dimensional space is a common and productive first step. For example, an autoencoder[13,14] (AE) maps each input spectrum to a point in a latent space of low dimensionality while simultaneously training a decoder to perform the inverse mapping of a point in the latent space back to a spectrum. However, the latent space variables do not inherently have a physical interpretation.[9,10,15–19] Specifically for the structure-property relationship problem, the data-driven discovery of correlative structure descriptors remains largely unsolved.

X-ray absorption spectroscopy (XAS) illustrates the complexity of structure-spectrum relationships. XAS is a widely used technique for materials characterization[20,21] due to its element specificity and its sensitivity to the local chemical environment of the absorbing atom.[22] Furthermore, modern synchrotron facilities enable high-throughput experimentation and in situ materials discovery methodologies.[23] In XAS experiments, a core electron is excited by an incident photon to empty states. The X-ray absorption spectrum exhibits steps, called edges, at clearly distinguishable energies where the absorption rises sharply. These are classified into K-, L-, and M-edges corresponding to $n$=1, 2, and 3, where $n$ is the principal quantum number of the core electron. For a given absorbing element, the XAS spectrum is typically divided into two regions relative to the edge. The region extending up to roughly 50 eV above the edge is referred to as X-ray Absorption Near Edge Structure (XANES), and beyond that is the Extended X-ray Absorption Fine Structure (EXAFS).[20]

XANES and EXAFS encode different types of information. EXAFS contains information about the radial distribution of first shell neighbors, and potentially second and third shell neighbors if the single-to-noise ratio is high enough. The analysis of EXAFS is well developed, with robust approaches available for extracting structural information.[24,25] In the framework of multiple scattering theory[26] (see Figure 1a), the EXAFS signal is dominated by the sum of direct back-scattering terms. In contrast, XANES depends on the interference of a more diverse set of interactions, such as the three-body path illustrated in Figure 1a. Furthermore, it contains rich information about the electronic structure, specifically dipole-allowed transitions to low-lying empty states. As a result, XANES spectral features can be correlated with local physical and structure descriptors, such as oxidation state, coordination number, and local symmetry.[20,21,26,27] However, XANES is more difficult to analyze than EXAFS. An in-depth analysis typically requires details of the electronic states of the material as well as an accurate physical model of the many-body core-hole effects.[26–29] Because of this inherent complexity, it is challenging to unravel the structure-spectrum relationships in the analysis of XANES.

Structure-spectrum relationships enable the extraction of structural information from XANES spectra measured on new materials, a core application of X-ray measurement in science and engineering. Consequently, the development of methods to infer structure descriptors from spectra is a central research problem in XANES analysis.[21,30–35] In contrast to the well-posed problem of computing a spectrum from a given material structure, this is an "inverse problem." The inverse problem may not always have a unique solution. Multiple structures may be consistent with a measured spectrum. Solving the inverse problem relies on identifying a set of robust structure descriptors that correlate to distinct spectral features or fingerprints. The search for such relationships has inspired decades of research in X-ray spectroscopy. In 1920, Bergengren discovered that the edge position strongly correlates with the oxidation state (OS) of the absorbing atom (see Figure 1b).[30] Since then, OS has been a common focus of the K-edge XANES analysis.[21,32,33,36] Other important structure descriptors include local symmetry (e.g., tetrahedral versus octahedral) and coordination number (CN) of the cation, which are correlated with the position and intensity of the pre-edge peak in early 3$d$ transition metal oxides (Figure 1c).[21,31,35,37] However, this empirical relationship shows strong system dependence and does not apply to late 3$d$ transition metals with vanishing pre-edge features, as illustrated in Figure 1d.[37–39] Beyond the first coordination shell, it is significantly more difficult to identify useful structure descriptors.

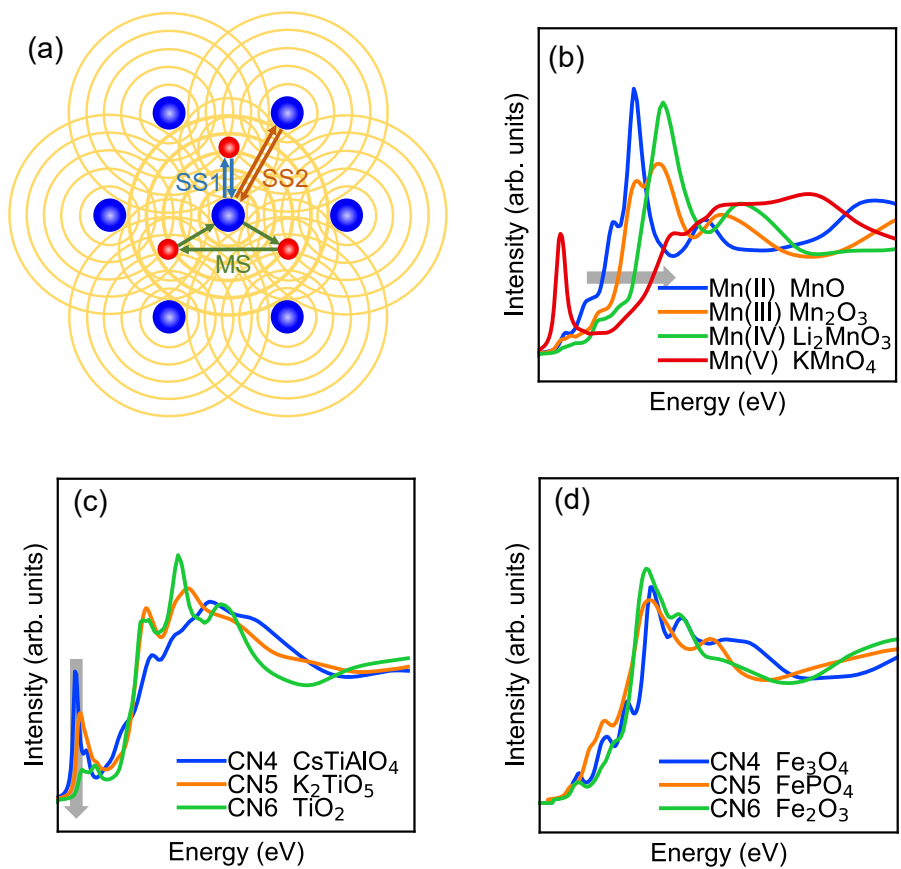

Figure 1. Real space scattering picture of X-ray absorption and exemplary spectral trends in K-edge XANES. (a) The final state electron amplitude on the central atom includes the sum of scattering contributions from all neighboring atoms. Scattering processes include: SS1 - single scattering from the first coordination shell, SS2 - single scattering from the second coordination shell, and MS – an exemplary multiple scattering path. (b) Simulated Mn K-edge XANES for a series of crystals in which the oxidation state of Mn varies showing the shift of the main edge (gray arrow). (c) Simulated Ti K-edge XANES for three oxide crystals with varying titanium coordination numbers showing the change in pre-edge feature intensity (gray arrow). CN4, CN5, and CN6 designate 4, 5, and 6-coordinated motifs. (d) Same for Fe K-edge XANES but showing no simple trend.

Despite the utility of existing structure descriptors (e.g., OS and CN), there is a strong demand to discover or engineer a sufficiently complete set of structure descriptors to support local structure prediction in complex materials spaces. The significance of a structure descriptor strongly depends on the underlying materials and the physical properties of interest. For example, the pre-peak features in transition metal K-edge XANES are sensitive to the distortion of the cation octahedral cage caused by substrate[40] or pressure.[41] Based on studies of Ti K-edge XANES and Li K-edge electron energy loss spectra, local distortions are identified as structure descriptors that play an important role in understanding the phase transformation and the fast lithium ion transport in lithium titanate.[42–44] To capture the spectral trends of nanostructures, bond-length-based structure descriptors have been investigated, which include metal-metal bond lengths in metallic (Pd K-edge)[18] and bimetallic nanoclusters (Pd K-edge and Au $L_3$-edge),[45] average Fe-O bond length in Fe oxide clusters (Fe K-edge),[34] and Co-C and Co-O bond lengths in a Co single atom catalyst complex (Co K-edge).[46] In addition, the chemical composition (e.g., hydrogen content) can also serve as a good descriptor for Pd nanoparticle catalysts exposed to $H_2$.[18] These examples illustrate the material specificity of the problem. Despite extensive studies,[32,33,47,48] a robust and widely applicable approach for identifying structure descriptors still does not exist.

Intuitively, one can label a spectral dataset with various structure descriptors and use visual inspection to search for qualitative trends. This empirical approach can succeed when there is an obvious trend, such as the correlation of edge shift to OS in Mn K-edge XANES (Figure 1b) or the correlation of pre-edge intensity to CN in Ti K-edge XANES (Figure 1c). However, this approach fails when the spectral trend is too complex to identify by visual inspection, e.g., the trend of CN in Fe K-edge XANES (Figure 1d). Nonetheless, in the latter situation, there is great potential to use data analytics tools to extract statistically meaningful trends.[31,33,38,48–52]

In this study, we focus on methods capable of learning a latent space that simultaneously represents spectra and correlates with structure descriptors. The resulting map is an interpretable, multi-dimensional structure-spectrum relationship. We identify key technical challenges that must be met to achieve this goal and develop new data analytics tools to address them. Specifically, we build on the adversarial autoencoder (AAE)[53] by adding a new rank constraint that drives each latent space variable to track a target structure descriptor. Taken together, our rank-constrained adversarial autoencoder (RankAAE) method captures spectral variations along latent space dimensions and correlates them with physically interpretable structure descriptors. We show that the method works across a multidimensional latent space incorporating a set of structure descriptors in a single training procedure.

To demonstrate the RankAAE method, we apply it to transition metal cation K-edge XANES from eight 3*d* transition metal oxide families (Ti-Cu). A large database of simulated spectra is calculated for crystal structures drawn from the Materials Project database, which exhibits diverse local cation chemical environments. This represents a broad coverage of materials compositions, including binary, ternary and quaternary oxides. The database includes over fifty thousand individual cation spectra. With this database, we use the RankAAE methodology to build structure-spectrum relationships and characterize the effectiveness of the method. We show how it can be used by a domain expert to explore candidate structure descriptors and their corresponding spectral trends. In particular, this method can be used to compare the performance of different combinations of structure descriptors in multi-dimensional models for structure-spectrum relationships. For the transition metal oxides under study, the RankAAE method enabled us to both recover historical structure-spectrum relationships[21,30,31,35–37,54] and to reveal new ones hidden in the data.

We henceforth summarize the structure of our manuscript. Latent space methods are briefly reviewed in Section II. Section III presents the main results demonstrating the use of the RankAAE method. The results are discussed in Section IV, including the physical interpretation of the emerging trends. Conclusions appear in Section V. Technical details of the methodology are described in Section VI.

## II. LATENT SPACE METHODS

Dimensionality reduction techniques pertinent to the structure-spectrum problem are briefly described here. In particular, we place our RankAAE method in the context of previous work.

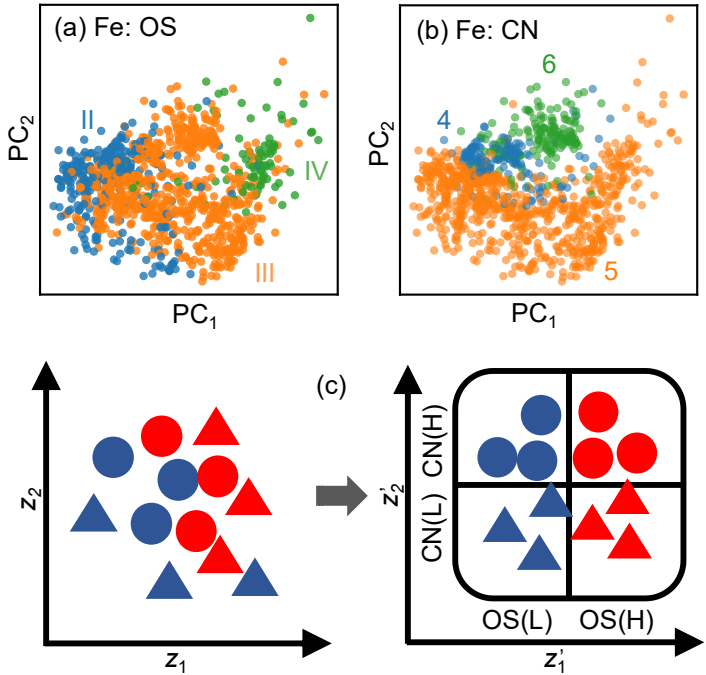

Figure 2. Dimensionality reduction with correlation to structure descriptors. (a, b) Principal component analysis of simulated Fe K-edge XANES spectra. Data points are colored according to the indicated structure descriptors: (a) Fe oxidation state (OS); (b) Fe coordination number (CN). (c) Schematic illustration of feature entanglement in the spectrum latent space $\{z_1, z_2\}$ (left) and the goal of transforming the latent space into $\{z'_1, z'_2\}$ (right), where $z'_1$ and $z'_2$ align with structure descriptors OS and CN, respectively. OS is represented by color (blue: low OS; red: high OS), and CN is represented by shape (triangle: low CN; circle: high CN).

One example of a simple, linear method commonly used to extract trends in data is principal component analysis (PCA). Liu *et al.* applied PCA to the Cu K-edge of $Cu_xPd_y$ bimetallic nanoparticles and identified patterns correlated with $Cu_xPd_y$ cluster types.[55] Carbone *et al.* performed PCA on K-edge XANES of eight $3d$ transition metal families and identified clear patterns, where the data points in the reduced dimensional space are distributed into identifiable clusters ordered according to the label of the absorbing cation CN=4, 5, and 6.[38] Similar patterns are shown here for Fe OS and CN in Figure 2(a, b), where the simulated spectra were drawn from the database used in the present study (see Methods). Fe OS exhibits a meaningful pattern in the reduced-dimensional space, where the gradient of the OS is roughly along the horizontal direction. Similarly, the ordered PCA pattern of Fe CN is roughly along the vertical direction, in stark contrast to the vague trend in the raw spectra of Figure 1d. The PCA decomposition of Fe K-edge XANES spectra in Figure 2 provides strong evidence that OS and CN are good structure descriptors. However, the variation of OS or CN is highly non-linear with respect to the PCA axes, and the patterns of the two descriptors are thus intertwined, as illustrated schematically in Figure 2c (left). This example illustrates the limitations of standard linear dimensionality reduction methods, such as PCA, for quantitative structure-spectrum mapping.

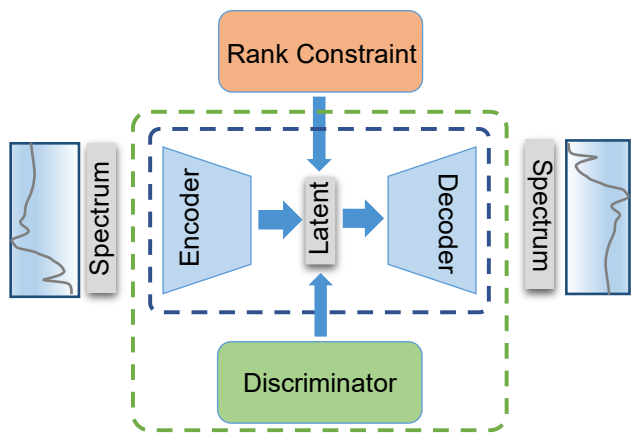

Figure 3. Network structure of the RankAAE method. A standard autoencoder (blue dashed rectangle) creates a latent space representation of the data that faithfully reproduces the spectra in the training dataset. An adversarial autoencoder (green dashed rectangle) introduces a discriminator on top of the regular autoencoder that regularizes the latent space. RankAAE further adds a rank constraint (orange) to the latent space to establish correlations between specific latent space dimensions and target physical descriptors.

A more promising dimensionality reduction approach uses autoencoder (AE)-based methods.[13,14] An AE generally consists of two neural networks trained in tandem to approximate the identity function: an encoder that compresses vector data sets to a lower dimensional latent space and a decoder that reverses this mapping (see the blue dashed box in Figure 3). An AE can perform sophisticated non-linear data compression and capture statistically relevant information in the latent space. For example, Routh *et al.* trained an autoencoder to transform simulated Pd K-edge XANES spectra of small Pd clusters to a latent representation and revealed a strong correlation between structure descriptors (CN, interatomic distance, and hydrogen content) and latent space variables.[18] However, in studies that require continuous sampling of the latent space, the mapping created by a basic AE can be problematic. For example, when applied to image reconstruction problems, some regions of the latent space do not decode to sensible images.[56] In a chemical science example, the latent space of an AE trained on SMILES encodings of molecules[57] exhibited "dead areas" that did not decode to valid molecules.[17] In the present work, the spectrum reconstruction from the decoder can yield unphysical spectra (Figure S1a). These failures are due to the lack of regularization of the latent space. Some points in the latent space may not correspond to regions in the neighborhood of any previously seen training data, causing the decoder to produce unpredictable results. Variational (VAEs),[10,17,58] Wasserstein (WAE)[9] and adversarial autoencoders (AAEs, green dashed box in Figure 3)[53,59,60] tackle this problem by regularizing the latent space during training, forcing the training data coverage of the latent space to be more complete. Although different in technical detail, all result in models capable of performing robust data generation (i.e., generative models). The latent space can then be sampled continuously, with decoded signals that remain valid in the target application, e.g., yielding physical spectra. For example, a VAE has been used to analyze spectral functions,[19] and VAEs, WAE and AAEs all have been used as generative models to search for new molecules/materials from the latent space.[9,17,59,60]

Robust, data-driven methodologies to uncover structure-spectrum relationships and discover new descriptors must address three key technical challenges:

1. *Spectrum validity*: a data-driven method is needed to construct a set of latent variables that serves as a proxy for the spectrum. In practice, the latent space needs to be regularized to ensure that all data points in the latent space correspond to physical spectra.
2. *Structure mapping*: the method needs to establish quantitative mappings between the latent space and structure descriptors, such that the statistical importance of various descriptors can be assessed and compared.
3. *Feature disentanglement*: the method must disentangle spectral contributions driven by different underlying structural and chemical characteristics.

For example, in the case of XANES, the spectral variation reflects the net effects from multiple sources that are often convoluted in overlapping energy ranges. Even with prior knowledge from domain-specific expertise, it is currently impossible to disentangle the impact of multiple structure descriptors. To achieve disentanglement, the latent space needs to be further organized, such that each latent space dimension aligns with a specific structure descriptor. Currently, none of the off-the-shelf data analytics tools simultaneously satisfies these three requirements of *spectrum validity, structure mapping,* and *feature disentanglement*.

New methods need to be developed to simultaneously address all three challenges. We focus on VAE/AAE methods. Because they are generative models in nature, they already satisfy the *spectrum validity* requirement. Joint training of a property prediction network with a VAE can address the *structure mapping* requirement,[17] however it does not solve the *feature disentanglement* problem. Specifically, even with the addition of joint training, each latent variable does not represent a target property directly since it can be entangled with multiple physical descriptors. To construct a direct latent variable-structure descriptor mapping, the latent space of a VAE/AAE needs to be reorganized to align each dimension with a structure descriptor. Figure 2c (right) depicts this idea: the goal is to have the latent variable $z_1'$ only depend on OS (symbol color) monotonically, while $z_2'$ only depends on CN (symbol shape) monotonically. To accomplish this, additional constraints need to be engineered and applied to a VAE or AAE.

The RankAAE method, developed in this work, accomplishes this goal with a novel "rank" constraint specifically designed to organize the latent space in the way depicted in Figure 2c. The implementation balances the goal of aligning each latent space variable with a target structure descriptor while minimizing the impact on the other, statistical characteristics of the latent space representation learned with the AAE. The technical details are described in Methods. Applied to a multi-dimensional latent space, the interplay of different structure descriptors is disentangled and the RankAAE method associates each dimension of the latent space with a well-defined structure descriptor.

## III. RESULTS

For our database of transition metal oxides, we simulate the XANES spectra for each symmetrically non-equivalent transition metal absorber using the multiple-scattering code,

FEFF9.[26,61] Details of the database development and computational methods are given in Methods.

Since XANES cation K-edge spectra probe local chemical environments, we investigate the structure descriptors that encode them. For the transition metal oxides, the cation-oxygen network characteristics are of particular importance (Figure 4a), although most crystal structures in the database also incorporate counterions that influence key chemical and structural attributes of the local transition metal and its environment. For simplicity, we restrict our attention to sites where the nearest neighbor shell contains only oxygen. In addition to the two well-known descriptors (OS and CN), we consider several other descriptors directed at capturing distortions in the nearest-neighbor shell of the absorbing cation and the influence of the second nearest-neighbor shell. The considered descriptors are detailed in Methods, but we outline them here briefly.

1) The average of the coordination numbers for the nearest-neighbor oxygen atoms (OCN),
2) The second-nearest-neighbor coordination number (CN-2),
3) The spread in the cation-oxygen bond lengths expressed as the nearest-neighbor radial standard deviation (NNRS),
4) The spread in the cation centered bond angles expressed as the nearest-neighbor angular standard deviation (NNAS),
5) The minimum oxygen-oxygen distance on the edges of the nearest neighbor polyhedron (MOOD), and
6) The point group symmetry order (PGSO).

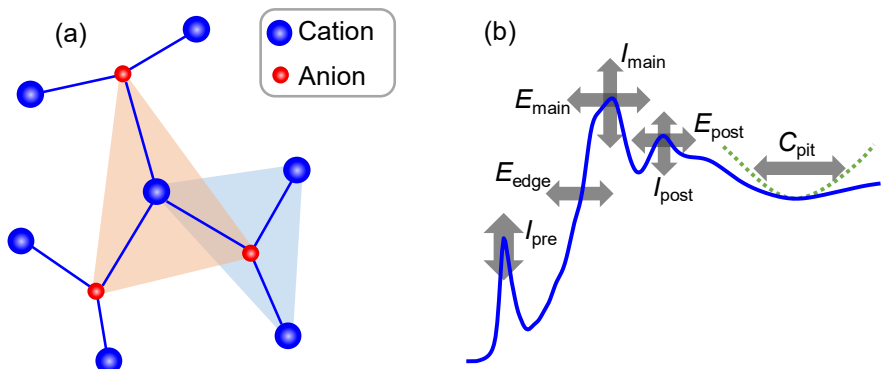

Figure 4. Structure and XANES spectral descriptors. (a) Schematic local structure around the central cation emphasizing the cation-oxygen network. Shaded areas are first coordination shells of cation (light red triangle) and oxygen (light blue triangle). (b) Schematic spectral descriptors applicable to typical cation K-edge XANES spectra in oxides: $E_{edge}$ – edge position, $I_{pre}$ – pre-peak intensity, $I_{main}$ – main peak intensity, $C_{pit}$ – post edge curvature, $E_{main}$ – main peak position, $E_{post}$ – post edge position, $I_{post}$ – post edge intensity.

To facilitate the discussion of structure-spectrum relationships, we adopt a set of spectral descriptors that capture the main spectral characteristics seen in a typical transition metal oxide XANES K-edge spectrum (Figure 4b). Described in more detail in Methods, these basic metrics will be referenced throughout our work.

We present our results in two parts. First, we illustrate the use of the RankAAE method and our validation of its performance for one family of materials, Vanadium (V) oxides. Then, we illustrate the performance of the method across the full set of transition metal oxides considered in this study.

## A. Application and Validation of RankAAE for Vanadium Oxides

The full scope of the V K-edge XANES data across the present vanadium oxide database is shown in Figure 5a. To illustrate the structure-spectrum relationships in this raw data, the spectra are colored according to the values of five structure descriptors in sequence from the top of the figure: OS, CN, OCN, NNRS, and MOOD, respectively. Visual inspection of the examples in Figure 5a shows color concentrations indicative of correlations between structure descriptors and spectral features, as one expects on physical grounds. For example, the position of the edge and the main peak ($E_{edge}$ and $E_{main}$) shift with OS, in line with the trend observed in experimental spectra.[21,31,35,37] Pre-edge peak intensity ($I_{pre}$) and main peak position and intensity ($E_{main}$ and $I_{main}$) change with CN, but $E_{main}$ and $I_{main}$ are also affected by OCN and MOOD. Higher energy features above the main peak show similar characteristics. This illustrates the entanglement of the contributions from different structure descriptors to trends in spectral features. This diverse dataset exemplifies the challenges that domain experts face when trying to expand the scope of structure descriptors in the analysis. While the overall trends in spectral features for OS and CN follow prior domain experience and physical models, the trends for additional spectral descriptors are too ambiguous or complicated to draw clear conclusions. A specific structure descriptor correlates with spectral features in several locations across the spectrum and each of those features can be affected by multiple structure descriptors. Taken together trends are obscured and physical interpretation is non-trivial.

We illustrate the use of the RankAAE method (Figure 3, Methods) through application to this vanadium oxide dataset. The RankAAE model is trained to map the spectra in the database to a latent space (center of Figure 3) of chosen dimensionality $N$ through an autoencoding procedure. For most of the results presented in this study we take $N = 6$. As a special type of AAE, the latent space resulting from the trained RankAAE model is regularized and each reconstruction maps to a physical spectrum (Figure S1b). In contrast, a standard AE model can yield wider variance and unphysical features in the spectra, such as excessive noise and even dramatic, sharp spikes (Figure S1a). In addition, the rank constraint is enforced for selected latent space variables to align with chosen structure descriptors, as described in Methods. Unconstrained dimensions represent residual characteristics of the spectral data set beyond the constrained dimensions. Training is repeated from scratch for each set of chosen structure descriptors. The details of dataset splitting (training, validation, and test) are listed in Table S1.

The regularized and aligned latent space resulting from the RankAAE method allows us to present spectral trends in an easily interpretable fashion. Reconstructed spectra are mapped from the latent space with the trained decoder. Once the model is trained, each point in the latent space decodes to a spectrum. In addition, due to our training procedure, each point should correlate to a set of structure descriptor values according to the imposed constraints. If the method is successful, the latent space will encode structure-spectrum relationships. By tracing a path through the latent space, we sample that mapping, tracking the correlated structure descriptor values together with the associated reconstructed spectra. In the simplest version of this picture, we can track the evolution of spectra along each axis of the latent space while holding the other latent space values constant, e.g., equal to zero (Figure S2a). With the correlation to target structure descriptors, this can isolate specific spectral features that correlate to that structure descriptor.

Given the complexity of the dataset and the goal of quantifying the structure-spectrum relationship in a multidimensional descriptor space, we want to sample the latent space more systematically, not just along isolated, one-dimensional paths. To this end, for each latent variable $z_i$, we average over spectra corresponding to different values of $\{z_j\}_{j \neq i}$. In this way, the evolution of spectra along direction $z_i$ is statistically representative of the full dataset. Methods section describes this averaging procedure. Figures S2a and S2b compare the sampling along each isolated axis with the averaging procedure. The two approaches result in consistent trends. We adopt the averaging procedure in this work for its statistical rigor.

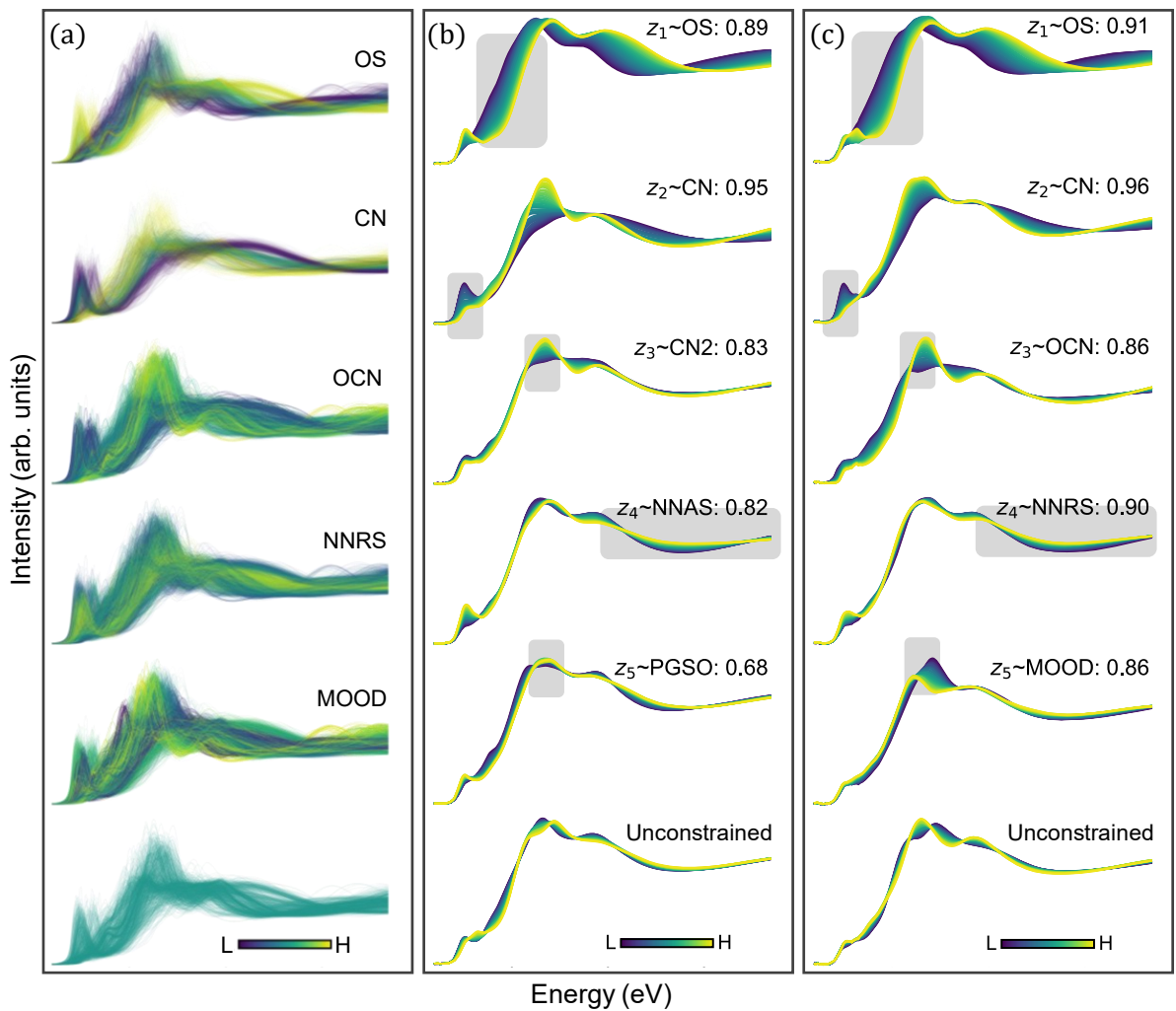

Figure 5. Correlating structure descriptors to spectral trends for V K-edge XANES from vanadium oxides. (a) Simulated spectra of vanadium oxides colored according to the value of specific structure descriptors: OS, CN, OCN, NNRS, MOOD, and no designation. (b, c) Reconstructed spectra from trained RankAAE models. As described in the text, each subplot maps the spectral trend associated with one of the six dimensions in the latent space. Spectra are colored by the value of the corresponding latent variable. Five of the latent space variables are constrained to correlate with target structure descriptors while the sixth variable is unconstrained. In (b), an initial set of structure descriptors is chosen for training: OS, CN, CN-2, NNAS, and PGSO. In (c) a final set of structure descriptors is chosen: OS, CN, OCN, NNRS, and MOOD. For each descriptor, the degree of correlation to the corresponding latent space variable is quantified by an F1 score (for OS and CN) or the Spearman rank correlation coefficient[62] (for the other descriptors), annotated next to each plot. The shaded area marks the location of the primary spectral variation characteristic of the spectral trend.

For the vanadium oxide dataset, Figures 5b and 5c show a series of trends for two trained RankAAE models respectively, each with five constrained latent space dimensions and one unconstrained dimension. Each subpanel shows the reconstructed spectral evolution along one dimension of the latent space. The smooth variation in spectra highlights the success in regularizing the latent space. The distinctive variations from panel to panel provide the input for interpreting the trends in terms of the target structure descriptors and assessing their relative utility.

It is not known *a priori* which structure descriptors capture the most significant spectral variation given a diverse dataset of materials structures. Figure 5b shows our "initial guess" for these structure descriptors based on prior domain knowledge. We evaluate the results according to several criteria. First, we monitor the quantitative correlation between the latent space variable and the target descriptor as detailed below. Second, we look for emerging spectral trends in distinct regions of the XANES spectrum. To be most useful for applications, those trends should be distinguishable. Finally, we monitor the amplitude of the spectral variation captured by the unconstrained latent space variable $z_6$. In essence, the last metric captures the spectral variance outside the scope of that driven by the target structure descriptors. Reducing it should indicate improvements in the completeness of the descriptor set.

For the results in Figure 5b, we see that the target descriptors are captured with generally high correlation (inset values in each panel). Furthermore, for each descriptor, a primary trend in the spectra can be identified (shaded area). In addition, there are other spectral trends associated with each descriptor across the energy range. For example, CN is correlated with both pre-edge intensity and main peak intensity. Thus, while it is convenient to isolate localized spectral features for discussion, each structure descriptor is associated with extended fingerprints with contributions from spectral features across the full energy range. We also note an overlap in the trends. For example, OS, CN, and NNAS all correlate to pre-edge intensity. Finally, the unconstrained dimension exhibits a relatively small spectral variation.

The process of refining the choices of structure descriptors is iterative and still requires a "human in the loop." As an example, we retain the well-known OS and CN descriptors that clearly show a strong, systematic spectral trend and consider alternatives for other structure descriptors. Figure S3 shows the change in spectral trend as each of those structure descriptors is replaced one by one. The OCN descriptor improves over CN-2 with a somewhat higher correlation to the latent space variable and qualitatively supports a stronger spectral trend in the shape of the primary peak. Introducing NNRS retains the correlation to the pit curvature while reducing the impact on pre-edge peak intensity. The MOOD descriptor exhibits a much-improved correlation to the latent space variable compared to PSGO (0.86 versus 0.68). In this example, the spectral variation attributable to the unconstrained latent space dimension remains relatively small. The spectral trend for the final structure descriptor set is presented in Figure 5c. In comparing the trends with the initial result in Figure 5b, the interaction between the descriptor choices can be seen. The pre-edge intensity trend for OS has been suppressed, and the spectral trend around the main peak shape for CN has been concentrated. Also, some of the overlaps in the spectral trends have been reduced. In particular, the pre-edge peak intensity is now more specifically correlated to the CN descriptor, and to a smaller degree to NNRS. This set is likely not unique, but it captures most of the spectral variance in the vanadium oxide database.

A different perspective on descriptor development is to consider how the model evolves as descriptors are added one by one. This is illustrated in Figure S4 for the five descriptors used in

Figure 5c. The quantitative correlation between each structure descriptor and the corresponding latent space variable is stable. The qualitative spectral trends are also similar, although we observe a clear sharpening of them. For example, the initial pre-edge feature in the OS trend systematically reduces to essentially zero. The CN trend consolidates into clear pre-edge and main peak features. At the same time, the OCN trend is capturing main peak shape changes. Finally, we also see the stepwise reduction in the amplitude of the unconstrained latent variables as additional structure descriptors are added. This strongly supports the notion that each new structure descriptor is extending the model to capture additional variation in the spectra. It also illustrates how the RankAAE approach overcomes the complexity of intertwined spectra trends in the raw spectral data (Figure 5a). As the set of structure descriptors incorporated into the trained model is modified, a set of distinguishable spectra trends emerges.

From the fully trained models, we can identify regions of the spectra with "primary spectral trends," each one linked to a specific latent variable $z_i$ and the associated structure descriptor (Figures 5b and 5c). Other significant concentrations of spectral variations constitute "secondary spectral trends." Together, they represent a spectral fingerprint. For the specific model for structure-spectrum relationships shown in Figure 5c, we identify the following trends:

*Oxidation state* (OS): The edge position shifts to higher energy with the increase of atomic charge due to shielding effects (Figure 5c, $z_1$ and Figure 4b, $E_{\text{edge}}$). In addition, there is a prominent secondary spectral trend: the higher energy portion of the spectrum also shifts horizontally (Figure 4b, $E_{\text{post}}$).

*Cation coordination number* (CN): The pre-edge peak intensity decreases sharply with the increase of CN (Figure 5c, $z_2$ and Figure 4b, $I_{\text{pre}}$). In addition, the intensity change after the main peak constitutes an identifiable secondary spectral trend (Figure 5b, $I_{\text{post}}$).

*Oxygen coordination number* (OCN): The intensity of the main peak increases with the increase of OCN (Figure 5c, $z_3$ and Figure 4b, $I_{\text{main}}$). Overall, OCN caused smaller but more focused changes than CN and alters the main peak shape.

*Standard deviation in the nearest neighbor bond length* (NNRS): The oscillation in absorption intensities becomes weaker as NNRS increases, especially at higher energies (Figure 5c, $z_4$ and Figure 4b, $C_{\text{pit}}$). In addition, there is also a small variation in the pre-edge peak. In general, this is a mild contribution to the total signal.

*Minimum oxygen-oxygen distance* (MOOD): This contribution to the total signal is even smaller than the other descriptors. However, the variation at the main peak is strong: the position of the main peak shifts to lower energy with increasing MOOD (Figure 5c, $z_5$ and Figure 4b, $E_{\text{main}}$).

*Unconstrained latent variable $z_6$*: By design, an AAE disentangles its latent space. In the case of the RankAAE model presented here, $z_6$ is disentangled from and contains information supplementary to the constrained variables $\{z_1, z_2, \ldots, z_5\}$. In other words, the plot for $z_6$ in Figure 5c represents *all other* spectral trends not captured by the five structure descriptors.

Having illustrated the characteristics of RankAAE models, we now describe quantitative validation. First, we characterize the correlation between target descriptors and latent space dimensions. To be clear, the value of a latent variable does not equal the value of the structure descriptor directly. However, each latent variable is nearly monotonically correlated to a specific structure descriptor, due to the soft constraint in our loss function. For the same model presented in Figure 5c, latent space values are plotted against calculated structure descriptor values in Figure 6 for the test set in scatter plots for each of the five descriptors. The correlations are visually apparent. They are quantified using the F1 score[63] for categorical variables (OS and CN) and the Spearman rank correlation coefficient[62] (SRCC) for continuous variables (OCN, NNRS, and MOOD). An F1 score ranges from 0 for poor classification performance to 1 for perfect classification performance. An SRCC value of 0 occurs when there is no correlation; a value of 1 indicates a perfect, monotonic, positive correlation. Furthermore, the degree of correlation depends on the choice of descriptor, as illustrated in Figure S3. As shown by the annotations in Figure 5c and summarized in Figure S5a, all the F1 scores are above 0.91 and all the SRCCs are above 0.86, highlighting the ability of the RankAAE methodology to drive latent variables to track structure descriptors in a nearly monotonic fashion.

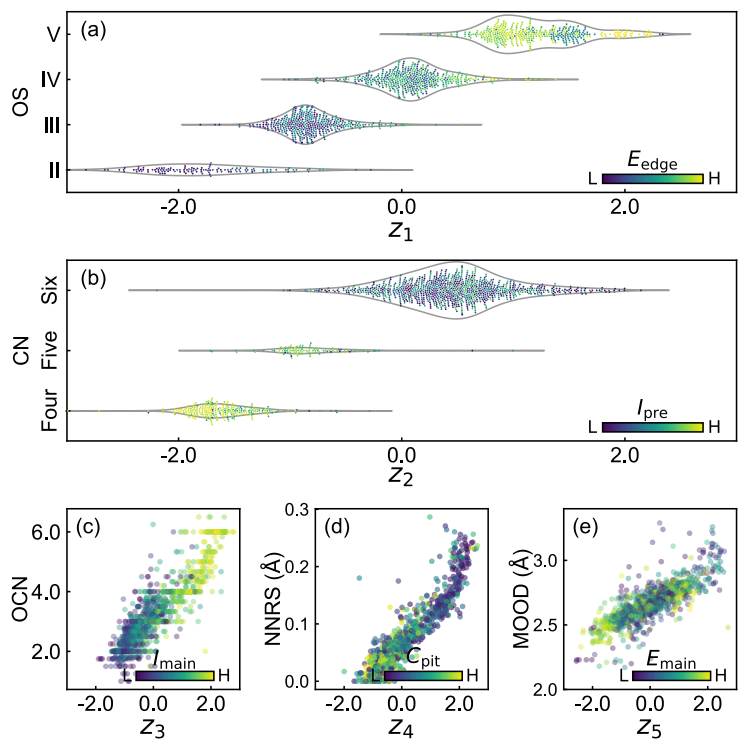

Figure 6. Correlation between the RankAAE latent space variable (x-axis), target structure descriptor (y-axis), and emergent spectrum descriptor (color) for the vanadium oxide data set. Violin plots are shown for categorical variables OS (a) and CN (b). Scatter plots are shown for continuous variables OCN (c), NNRS (d), and MOOD (e).

Next, the emergent spectral trends encoded in the reconstructed spectra in the RankAAE model in Figure 5c are compared to features in the ground truth spectra, i.e., the spectra directly from the data set. For simplicity, we focus on the primary spectral trends that have been identified in Figure 5c. We compute these primary spectral descriptors for each of the underlying spectra in the test set and present them in Figure 6. Visual inspection verifies that the evolution of the spectral descriptors tracks both the latent variables and the structure descriptors. To quantify these relationships, the F1 scores and SRCCs between the spectral descriptor in the data set and the latent space variable are computed (Figure S5b). The coefficients are smaller than those for the structure descriptors (Figure S5a). However, they are still significant, especially considering that only one local spectra feature has been used. Also, the additional fluctuations associated with the spectral features as emergent in a multi-dimensional space play a role. Finally, the reduction of the correlation analysis to single dimensions is an oversimplification. For example, for $z_5$, the correlation with spectral descriptor ($E_{main}$) is only clear after categorizing the data points according to the CN because the main peak variation is also affected by CN. This effect is shown in Figure S6 using the test portion of the vanadium oxide dataset as an example. The data points aggregate to a strip for each CN category. Inside each strip, the correlation between $z_5$ and $E_{main}$ is much clearer, especially for CN5 and CN6, than when considering all the data in aggregate. This illustrates that a local spectral feature, such as $E_{main}$, can still be affected by multiple structure descriptors.

Taken together, these validation results demonstrate that the trained RankAAE model provides multidimensional, physical structure-spectrum relationships for the vanadium oxide dataset.

## B. RankAAE Performance across the Full Set of Transition Metal Oxides

The characteristics, trends, validation, and the final set of five structure descriptors shown for the specific example of V oxides carry over to the full set of 3$d$ transition metal families considered in this work. The spectral data colored by the structure descriptors reveals trends with OS and CN, but ambiguity remains when considering the larger set of descriptors (Figure S7). RankAAE models are trained on the datasets of Ti, Cr, V, Mn, Fe, Co, Ni, and Cu oxides. An independent model is created for each material family. The same set of structure descriptors is used for the entire series to facilitate comparisons. The reconstructed spectra versus each latent space dimension exhibit smooth trends (Figure S8). The amplitude of spectral variation for the unconstrained latent space variable $z_6$ varies across the series. This suggests an opportunity to further fine-tune the choice of structure descriptors. Examining the spectral trends, the same set of spectral features can be tracked, although the specific energy ranges (relative to the edge) and the relative importance of the primary versus secondary vary.

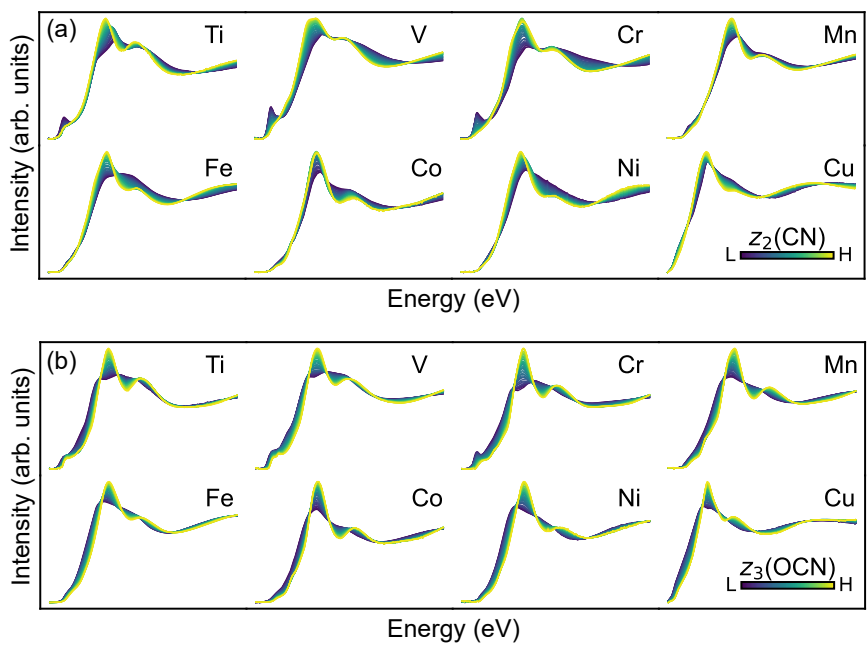

Figure 7. RankAAE derived spectrum variation trends for CN (a) and OCN (b) over Ti, V, Cr, Mn, Fe, Co, Ni, and Cu oxides data sets. The spectra are colored by underlying latent variables $z_2$ and $z_3$ targeting CN and OCN, respectively.

To illustrate these trends further, Figure 7a shows the trend of the spectral variations for CN across the series of first-row transition metals. The RankAAE models reveal two distinct primary trends. First, like for V oxides, the pre-edge peak variation is the primary spectral trend for Ti, Cr, and Mn oxides. The pre-edge peak intensity is strongest for V and Cr, intermediate for Ti, and small, but discernable for Mn. Second, the late transition metals (Fe, Co, Ni, and Cu) exhibit minimal intensity in the pre-edge peak region of the XANES K-edge spectrum.[37,38] For these metals, the secondary spectral trend overweighs the primary trend: the second peak intensity above the main edge, which is at least 5 eV away from the main peak (Figure 4b, $I_{post}$), correlates with CN.

As a second example, the spectral trend for OCN is shown in Figure 7b. There is a universal pattern: spectral intensity changes at the main peak. For Cu oxides, the intensity variation is localized to that around the main peak. It is also dominant for Mn oxides, Co oxides and Ni oxides. Other cases show more extensive, secondary spectral variation. Overall, the spectral features associated with OCN are distinguishable from those driven by CN. However, Cr oxide is an exception, where the trend associated with OCN is very similar to the trend for CN.

Validation of all the RankAAE models using the test sets for each material family shows strong correlations between latent space values and the target structure descriptors as well as the identified spectral features (Figure S9). Quantitatively the F1 scores and SRCCs for structure descriptor to latent space values are high, as shown as insets in Figure S8 for each material and descriptor, as well as in summary form in Figure S5a. Most F1 scores are above 0.90 with a minimum of 0.82. Most SRCCs are above 0.85 with a minimum of 0.80. This indicates that the latent space in the RankAAE models captures the changes in structure descriptors sufficiently across the entire set of 3$d$ transition metal families. OS and CN exhibit high correlations in all cases. The correlation for the three new structure descriptors varies across the metal series, but is very strong in general, showing their potential for motif characterization in XANES analysis. Quantitative validation against the primary spectral descriptors (Figure S5b) supports the qualitative results presented in Figure S9. Some specific cases do highlight the limitations of relying on a single spectral descriptor. For example, the low correlation (SRCC=0.13) for the Mn oxide CN descriptor with the pre-edge intensity spectral descriptor traces to the relatively low intensity of the pre-edge feature, as noted above. However, the SRCC increases to 0.44 if $I_{post}$ is used. Consistently, $I_{post}$ captures CN better as a spectral descriptor for late transition metals (Mn-Cu).

## IV. DISCUSSION

The RankAAE model described here creates a multidimensional, continuous spectral representation that maps a target set of structure descriptors to XANES K-edge spectral characteristics. The *spectrum validity* is a key prerequisite for its interpretability. A meaningful structure-spectrum relationship can only be established on a properly regularized latent space. Each point in the latent space must reconstruct to a physical XANES spectrum. As illustrated in the vanadium K-edge data set in Figure S1a, a basic AE model does not meet the *spectrum validity* criteria, as discussed in the Introduction. In fact, the AE model does reconstruct the spectra very well for latent space points close to areas well represented in the original training data set, evidenced by the dominance of "normal" spectra in Figure S1a. However, there are regions with values that fall in the valid range overall, but for which the reconstruction can become catastrophically poor. These latent space points fall where the model is interpolating in regions sparsely represented in the training original dataset.[56] In contrast, the latent space of the RankAAE model is regularized by the adversarial constraint, which enforces *spectrum validity*, as shown in Figure S1b, consistently produces physically meaningful data points.

Analysis of the reconstructed spectra along specific dimensions in the latent space of the RankAAE associated with each structure descriptor reveals well-defined spectral trends. For use by domain experts, it is essential that these trends are robust, and that there is a reasonable workflow to allow the expert to assess which choices of a structure descriptor are capturing the essential spectral variations, and that the trends are physically interpretable.

The workflow for developing multidimensional models demonstrated here exhibits stability in the spectral trends that emerge. We revisit the test sequence where we start with a single constrained latent space variable and one free dimension. We also add a latent space variable that is constrained to a pseudo-random number sequence as a structure descriptor (NOISE) to probe the limit of the noise level in deriving spectral trends. Thus, we start with $N = 3$, and add dimensions together with constraints in succession. The root of this chain is OS, followed by CN, OCN, NNRS, and finally MOOD. For the vanadium oxide dataset, following the same analysis as presented in Figure 5c, the appropriately averaged, reconstructed spectra along each dimension are shown in Figure S4. Several important points emerge. First, by visual inspection, the trends shown for each dimension remain qualitatively consistent as new dimensions and structure descriptors are added to the model. For example, for OS, the edge shift and horizontal translation remain stable with different combinations of structure descriptors, although the amplitude of the changes varies quantitatively. In particular, in the test using OS as the only structure descriptor, the pre-edge variation is not localized. Instead, it appears in two latent variables ($z_1$ and unconstrained $z_3$). With the addition of CN and more structure descriptors, it is localized to $z_2$. This reflects the resolution of entangled trends as further structure descriptors are added. Second, the effectiveness of the constraint in forcing the latent space variables to track the target descriptors is nearly independent of the dimensionality of the model, as shown by the F1 scores and SRCCs that quantify the correlation between latent space variables and target structure descriptors (Figure S4). Third, the noise spectral trend always appears in a very narrow range independent of dimensionality and the choices of constraint on other latent dimensions. Finally, the amplitude of the unconstrained spectral trend systematically goes down as new dimensions and structure descriptors are added to the model. This gives a qualitative guide indicating that each additional descriptor is capturing new spectral variation. With the full complement of five structure descriptors in the model, the spectral variation associated with the unconstrained variable is close to the noise level in vanadium oxides.

Another important aspect of the workflow for a domain expert is to distinguish between different choices of target descriptors. Several criteria have already been illustrated, including distinguishable spectral trends, quantitative correlations between target structure descriptors and constrained latent space variables, and the amplitude of the unguided spectral trend. Here we further analyze the degree to which each new descriptor results in a qualitatively distinguishable spectral trend. In order to capture the energy dependence of each trend when generating new spectra by traversing the latent space, we examine a "differential spectrum". We specifically represent the total dynamic range of the trend by taking the difference between the last and first spectrum corresponding to the $95^{th}$ percentile and $5^{th}$ percentile values in the range of the latent space variable. We revisit the comparison of trends for different structure descriptors presented in Figures 5 and S3 for vanadium oxides through the differential spectra shown in Figure S10. This representation of the trends also exhibits the qualitative stability of trends for OS and CN as other structure descriptors are changed. Further, as previously highlighted, the overall amplitude of the unguided trend is reduced as the choice of structure descriptors is refined. As the other structure descriptors are swapped out, the qualitative changes in the trends can be seen. For example, OCN more narrowly isolates the change in the main peak intensity while CN-2 is convolved with the shift of the main peak energy. Also, the trend for MOOD captures changes to the main peak shape, position, and intensity as compared to the relative spread-out trend for PGSO. The spectral trends are an emergent characteristic of the RankAAE models, so they are not constrained to be linearly independent. Nonetheless, in the workflow, the degree of overlap can be monitored through the intercorrelation measured by the Cosine similarity,[64] with the maximum values noted in each panel of Figure S10. Overall, as the workflow proceeds, the degree of overlap is reduced. In our final model, the CN descriptor and unconstrained dimension result in the most independent trends while OS, OCN, NNRS, and MOOD derived trends remain correlated with each other to a moderate degree by this measure. Finally, this analysis points in the direction of future investigations to refine spectral fingerprints beyond the simplification of the specific, localized spectral descriptors in Figure 4b.

Turning to physical interpretation, each of the primary and secondary spectral trends identified above with structure descriptors can be understood. Among all the structure-spectrum relationships, the correlation between OS and edge position is probably the most well-known.[30] Conceptually, it corresponds to a basic physical principle: the energy cost to excite a core electron increases as the cation becomes more positively charged. The secondary spectral trend, the correlated shift of the post-edge intensity to higher energy, corresponds to the famous "molecular ruler",[20] also referred to as Natoli's rule:[21] the local bond length is inversely linked to the position of a peak. Here the ionic radius changes inversely with the atomic charge for the $3d$ transition metal cation, driving the metal-oxygen bond length. These two empirical rules were proposed independently. The unified spectral trend for OS that emerges from the RankAAE models captures these correlated effects of the cation oxidation state.

The pre-edge peak is linked to the breaking of centro-symmetry. The onset of the K-edge spectrum involves low-energy empty states on the cation that are typically of $3d$ orbital character. The $s \rightarrow d$ transition is formally forbidden for an ideal, centro-symmetric octahedral motif. In lower coordination motifs, e.g., tetrahedral, as well as distorted atomic motifs, hybridization with cation $4p$ orbitals leads to dipole-allowed transitions.[37] The pre-edge spectral trends associated with CN and NNRS are a consequence of the breaking of centro-symmetry. This is consistent with the original observations of Farges *et al.*, and subsequent work.[31,37]

Drawing on the multiple scattering representation of the absorption process (Figure 1a), the number of scatterers is an important factor that affects the absorption intensity, while path lengths and interference effects determine the energy position where the absorption intensity changes. The changes in main peak position, intensity, and shape are a key part of the spectral trend for both CN and OCN (Figure 5c, $z_2$ and $z_3$). The overall increase of intensity with increasing coordination number tracks the number of scatterers. For CN, the sharpening of the main peak shape with more weight at lower energy as well as the shift of the post peak features to lower energy are further examples of Natoli's rule. As coordination increases, so do the metal-oxygen bond lengths. Correspondingly, the path lengths get longer.

The OCN trend (Figure 5c, $z_3$) modulates the intensity around the main peak and to a lesser extent the post-peak region. The OCN descriptor correlates with the number of atoms in the second coordination shell, and it covers a broader range of atoms that can contribute to three- and four-body scattering paths (Figure 1a). With an increase in the number of such paths induced by the increase of such atoms, interference effects are stronger. But, as compared to the CN trend, the energy location differs because of interatomic distances. OCN essentially counts the atoms in the second coordination shell while CN counts the first coordination shell (Figure 4a), and hence is related to larger interatomic distance. It results in a longer scattering path length (Figure 1a, SS2 versus SS1), longer wavelength, and consequently a lower energy photoelectron. As a result, the OCN induces spectral variation only at lower energy while CN also induces variations at higher energies (Figure 5c $z_3$ vs $z_2$). While OCN and second shell coordination number count the same set of atoms, it is worth noting that the definition of cutoff radius for the second coordination shell is ambiguous. In contrast, OCN is well-defined in established algorithms.[65] As a result, better latent space-structure descriptor correlation can be observed for OCN (Figure 5c $z_3$ versus Figure 5b $z_3$)

The NNRS descriptor measures the spread in the cation-oxygen bond lengths in the first coordination shell (Figure 4a). The scattering path lengths are spread out with larger NNRS. Correspondingly, there is less reinforcement of constructive and destructive interference among the paths as a function of photoelectron energy. This, in turn, reduces the definition of the peaks and valleys in the spectrum. Specifically, as the NNRS increases, the peak-to-valley height decreases. This is expressed, for example, in the local curvature in the post-peak spectral region, quantified in the $C_{pit}$ spectral feature (Figure 4b). The curvature is higher (deeper, better-defined valley) when NNRS is smaller (Figure 5c $z_4$). In the database and this analysis, a larger NNRS value represents crystal structures with low symmetry and a spread of local bond lengths. It is worth noting that $C_{pit}$ is a specific spectral descriptor that can be clearly discerned for most transition metals. Nonetheless, for specific metals, e.g., Mn in Figure S5b, the correlation between $C_{pit}$ and $z_4$ is low with an SRCC of 0.05. As suggested by the spectral trends in Figure S8, the curvature at the main peak correlates with $z_4$ better where the SRCC increases to 0.33.

The variation in the local angles between the cation-oxygen bonds is another measure of spread in low symmetry local motifs. It flows through to the path lengths for three-body scattering paths (Figure 1a). In our comparison of different specific structure descriptors, we found that MOOD, the distance minimum between two oxygens in the first coordination shell, was the most effective metric for the RankAAE models (Figure 4a) and can be related to the 3-body scattering path (Figure 1a, MS) in an X-ray absorption process. Based on a few examples in Figure S11, the bending of the two axial oxygens towards the equatorial plane is one of the most intuitive structure characteristics that the underlying latent variable $z_5$ tracks. The distortion associated with the bending is expected to change the 3-body scattering path length. The triangular path of MS is relatively long and leads to oscillation at relatively low energy. In the case of vanadium oxides, it is located around the main peak. With increasing oxygen-oxygen distance, the corresponding wavelength increases and pushes the main peak position to lower energy.

## V. CONCLUSION

The approximately fifty-seven thousand simulated XANES K-edge data considered in this study span eight 3$d$ transition metal oxide families and exemplify complex data sets that domain experts are highly motivated to study. Our novel rank constraint introduces physical interpretability into the latent space representation of data produced by an AAE. Specifically, latent variables are guided to track specific structure descriptors. From there, the reconstruction from different regions in the latent space produces spectral variations reflecting the underlying changes in structure descriptors. The correlations between structure descriptors and spectral trends are confirmed to be both qualitatively and quantitatively observed in the ground truth data.

For the specific example of XANES analysis, we have shown that our method significantly extends the scope of spectral features a domain expert can analyze systematically, from local metrics such as edge position to extended fingerprints that can encompass the full energy range of XANES spectra. Leveraging the prior knowledge of domain experts, these spectral trends can be used to gain further insight into features in materials structures and their relationship to spectral trends. Also, as illustrated by the set of new structure descriptors explored in this work, it is straightforward for a domain expert to explore new trends and new structure descriptors using the RankAAE method driven by their intuition.

From a machine-learning perspective, we also highlight the challenges associated with the extreme diversity of our datasets. First, the data is completely unbiased towards the specific methods used in this study. It consists of materials data compiled independently by contributors to the Materials Project over many years, driven by goals unrelated to the scope of this work. Second, the target signal, namely a specific transition metal K-edge XANES spectrum, is sampled from a set of diverse crystal structures with widely varying, albeit physically realistic, local structure motifs and associated atomic species. Consequently, the data encompasses a broad variety of physical complexities that affect the spectrum. These include different local chemical states of the transition metal cation, scattering amplitudes that differ among atomic species, and the distinct set of scattering paths (number of paths contributing, number of atoms in each path, and overall length of each path) associated with each cation structure motif. This means that our methods have numerically parsed a tremendous amount of diversity to identify the key trends we present. Thus, the successes that we demonstrate in this work are quite encouraging and highlight the potential of this, and related methods, for unearthing new spectral trends for impact on physical science and engineering.

In summary, we demonstrate the capability of RankAAE using the XANES structure-spectrum relationship as an example. However, the RankAAE is a *general* framework for enhancing the interpretability of an AAE. In principle, it is also applicable to other kinds of spectroscopic data, and in general, to *any* dataset in which one is attempting to discover physical correlations between target signals and associated driving characteristics.

# VI. METHODS

## A. Structure of RankAAE and Training

AAE,[53] VAE[58] and WAE[9] regularize the latent space to a target distribution. The VAE constrains its latent space by a KL divergence penalty to impose a prior distribution. The WAE performs similar regularization by a Wasserstein distance.[66] The AAE does not require a specified functional form of the penalty function to the prior distribution. It instead matches the aggregated posterior of the latent vector with the prior distribution by a discriminator via an adversarial training procedure. The AAE is reported to generate a better data manifold.[53,56,67,68] In this work, a Gaussian distribution is used. We have adopted the gradient reversal layer from the domain adversarial neural network to implement the adversarial training mechanism.[69]

The encoder, decoder, and discriminator (Figure 3) are all composed of 5 fully connected layers,

with a hidden layer size of 64. Both Dropout and Batch Normalization are used. A Parametric Rectified Linear Unit (PReLU)[70] function is added after every fully connected layer with two exceptions: (1) there is no activation function after the last layer in the encoder; (2) the last activation function in the decoder is Softplus.

We incorporate the mutual information regularization term from the Dual Adversarial AutoEncoder[71] to maximize the information shared between the latent space and the spectrum data. We have also used a smooth loss in the early stage of the training to speed up the convergence of network parameters.

With the performance evaluated on the validation set, the hyperparameters (learning rate, batch size, noise-based data augmentation, gradient reversal amplitude, hidden layer size, the number of hidden layers, and other aspects of the network structure) are optimized for vanadium oxides. The performance is balanced between the appearance of the spectral trends and the correlation of the latent variables with structure descriptors.

## B. Rank Constraint

A direct regularization of latent space using a structure descriptor value will inevitably distort the distribution. To avoid such distortion, we developed a novel regularization term based on the Kendall rank correlation coefficient (KRCC).[72] It enforces a monotonic dependence while being a soft enough constraint to allow variation in the final value. KRCC is built upon the concept of concordant and discordant pairs. In the application to RankAAE, a latent variable for spectrum $i$ can be denoted as $z_i$ and the corresponding structure descriptor can be denoted as $p_i$. Any pair of $(z_i, p_i)$ and $(z_j, p_j)$ are said to be concordant if both $z_i > z_j$ and $p_i > p_j$ or both $z_i < z_j$ and $p_i < p_j$; otherwise, they are said to be discordant. KRCC describes the monotonic dependence by the ratio of the number of concordant/discordant pairs. The counting is simplified using a sgn() function, i.e., sgn $(z_i - z_j)$ and sgn$(p_i - p_j)$. However, the gradient of the sgn() function is 0 except at 0, which is detrimental to the purpose of neural network parameter optimization. We have modified KRCC to a loss function with a proper gradient everywhere by removing sgn() function around $(z_i - z_j)$. The loss function $L_c(z)$ for each constrained dimension of the latent space is defined by:

$$f(z_i, z_j) = (z_i - z_j)\text{sgn}(p_i - p_j)$$

$$N_{con} = \sum_{i=1}^{n}\sum_{j=1}^{n} \Theta\big(f(z_i, z_j)\big)$$

$$N_{dis} = \sum_{i=1}^{n}\sum_{j=1}^{n} \Theta\big(-f(z_i, z_j)\big)$$

$$f_{con}(z) = \frac{1}{\max[N_{con}, N_{dis}]}\sum_{i=1}^{n}\sum_{j=1}^{n} \Theta\big(f(z_i, z_j)\big) f(z_i, z_j)$$

$$f_{dis}(z) = \frac{1}{N_{dis}}\sum_{i=1}^{n}\sum_{j=1}^{n} \Theta\big(-f(z_i, z_j)\big) f(z_i, z_j)$$

$$L_c(z) = -\frac{1}{n(n-1)} N_{\text{dis}} \left( f_{con}(z) + f_{dis}(z) \right) \tag{1}$$

where $z$ is a latent variable, $p$ is a structure descriptor, $n$ is the number of samples in a mini-batch in network training, $i$ and $j$ are sample indices in a minibatch, sgn() is a function to extract the sign of a real number, $\Theta()$ is a Heaviside step function being 1 for positive values and 0 for other values. Abbreviations "$_{\text{con}}$" and "$_{\text{dis}}$" designate concordant and discordant. $N_{\text{con}}$ and $N_{\text{dis}}$ represent the number of concordant and discordant pairs in the minibatch for the constrained dimension of the latent space. The minimum value of $N_{\text{dis},m}$ is clamped to 1 to avoid the singularity. The prefactors for the concordant and discordant pair contributions to the loss function are scaled to be comparable early in the training and then to have a diminishing impact as more data points become concordant. Overall, we find that this approach minimizes the effect of rank constraint on the distribution of the latent space variable.

## C. Data Acquisition and Preprocessing

The crystal structures are pulled from the Material Project.[54] Each crystal structure is a local minimum in energy with lattice parameters and atomic coordinates.[73] All the crystal structures containing each target transition metal atom and oxygen are included. The XANES K-Edge spectra are simulated using the FEFF9 program[61] for the resulting database of about forty thousand unique atomic sites identified in the oxides for the eight transition metals considered here with the condition that the first neighbor shell of the transition metal site only contains oxygen. The breakdown of the dataset according to transition metal is given in Table S1.

In this work, XANES is defined as the spectrum with energy ranging from the start of the edge of the spectrum to 50 eV above the edge. To create feature vectors that are directly comparable for each metal considered, the spectra are interpolated to a fixed energy grid, equally spaced with 256 points. For machine learning, each data set is partitioned with a random selection of 70%, 15%, and 15% of the spectra to form training, validation, and test sets, respectively. The size of the sets is given in Table S1 for each transition metal.

## D. Structure Descriptor Calculation

Eight structure descriptors have been explored in this study:

1) OS: oxidation state of a cation represented by an integer. OS is determined using the maximum a posteriori (MAP) estimation method in *pymatgen*.[74–76]
2) CN: coordination number in the first coordination shell around a specified transition metal cation. CN is determined using the CrystalNN algorithm in the *pymatgen* package.[65,75]
3) CN-2: coordination number in the second coordination shell for each target transition metal cation. CN-2 is estimated by the number of atoms that falls in the layer between radius $r_1$ and $r_2$, chosen for each transition metal based on the radial distribution of the atoms.
4) OCN: oxygen coordination number. OCN is computed as an average of the coordination numbers for the nearest neighbor oxygen atoms to the target transition metal cation. The coordination number of each oxygen atom is determined using the CrystalNN algorithm in the *pymatgen* package.[65,75]
5) NNRS: the standard deviation of the bond lengths from the target transition metal cation to the nearest neighbor oxygen atoms.

6) NNAS: the standard deviation of the bond angles with the cation as the vertex and two nearest neighbors as endpoints.
7) MOOD: the minimum oxygen-oxygen distance on the edges of the nearest neighbor polyhedron encompassing the target transition metal cation. The polyhedron is determined by the CrystalNN algorithm in the *pymatgen* package.[65,75]
8) PSGO: the point group symmetry order (PGSO) represented by an integer, computed as the total number of symmetry operations for the point group for the full crystal space group determined by the *pymatgen* package.[65,75]

## E. Spectrum Descriptor Calculation

The spectrum descriptors shown in Figure 4b are defined as:

1) $E_{edge}$: the position of the absorption edge.
2) $I_{pre}$: pre-peak intensity, the maximum intensity among the peaks occurring at an energy lower than the edge.
3) $E_{main}$: main peak position, the energy value at the first peak after the edge.
4) $I_{main}$: main peak intensity. For numerical stability purposes, we use the average over a 1 eV window centered at $E_{main}$.
5) $C_{pit}$: post edge curvature. The position of the pit is found by a local minimum of the spectrum in the portion at least 20 eV above the edge. The curvature is computed as the second order derivative. To avoid numerical instabilities, the curvature is averaged over a 10 eV window centered at the pit.
6) $E_{post}$: second peak position. A second peak is identified by a local maximum of the spectrum. To avoid ambiguities in peak counting, we considered $E_{edge}$ + 15 eV for transition metals Ti through Mn and $E_{edge}$ + 20 eV for Fe through Cu as a proxy for $E_{post}$. The numerical test in Figure S5b confirms the effectiveness of this approach.
7) $I_{post}$: absorption intensity at $E_{post}$.

## F. Spectral Variation Plot Generation

Specific latent space samples are fed to the decoder to generate reconstructed spectral variation sequences. For each latent variable, 50 points on an equally spaced grid are sampled. From the range of the latent space values encoded from the test set spectra, the latent space grid is chosen to extend from the 5$^{th}$ to the 95$^{th}$ percentile values. For Figure S2a, a spectrum is reconstructed for each of the 50 values of the specified latent space variable, setting the others equal to zero. For Figure S2b and all of the other reconstructed spectral trends, an averaging procedure is applied to sample the other latent space variables. Specifically, the average of reconstructed spectra for each value of the target latent space variable is carried out over 10000 samples drawn from a multivariant Gaussian distribution to fill the rest latent variables. There are 50 spectra in each spectral trend presented.

## G. Model Selection

For each dataset, we repeat the training 100 times with different random seeds for the neural network parameters, and all the discussions are based on the best one. The performance for each model is assessed by:

$$S = -\max_{i,j}|\rho_{ij}| + \sum_i |\rho'_i| \tag{2}$$

where $\rho_{ij}$ is the Spearman rank correlation coefficient (SRCC) between a guided latent variable $z_i$ and an unconstrained latent variable $z_j$, $\rho'_i$ is the correlation score (F1 score for OS and CN, SRCC for other structure descriptors) between a constrained latent variable $z_i$ and the target structure descriptor. To drive the performance metrics to have equal contributions, the z-scores are computed across the 100 models by subtracting the average and dividing the standard deviation for each $\rho_{ij}$ and $\rho'_i$. Then the z-scores[77] are used to evaluate $S$ in Eq. (2). A good model maximizes the latent-structure correlation and minimizes the intercorrelation between the unconstrained latent variables and the other dimensions. Therefore, a larger $S$ indicates better performance. For this study, it is sufficient to consider the two terms with equal weight. However, for other studies, the assessment of model performance can be further refined by scaling the two terms differently. Table S2 details the statistics of different loss terms (prior to z-score normalization) that enter Eq. (2), as well as the standard reconstruction error attributed to the autoencoder. For the current study, the reconstruction error is small and it does not vary significantly. Hence, we did not include reconstruction error in the model selection. The other terms vary within reasonable bounds. For further insight, the correlation plots for the worst model out of 100 runs by this criterion for the case of V oxide are shown in Figure S12. Encouragingly, the correlations for that model are still reasonably good.

## ACKNOWLEDGMENT

This research is based upon work supported by the U.S. Department of Energy, Office of Science, Office Basic Energy Sciences, under Award Number FWP PS-030. This research also used the Theory and Computation facility of the Center for Functional Nanomaterials (CFN), which is a U.S. Department of Energy Office of Science User Facility, at Brookhaven National Laboratory under Contract No. DE-SC0012704.

## SUPPORTING INFORMATION

Supporting Information is available.

# Table of Contents

    A new approach to harness powerful machine learning methods based on autoencoders results in compact, physically interpretable representations of complex spectral datasets. This paradigm shift enables discovery of structure-spectrum relationships, applicable to a wide range of scientific fields. A showcase study demonstrates new relationships to extract more structure information from X-ray absorption spectra.

Authors: *Zhu Liang*, *Matthew R. Carbone*, *Wei Chen*, *Fanchen Meng*, *Eli Stavitski*, *Deyu Lu,** *Mark S. Hybertsen,** and *Xiaohui Qu**

Title: Decoding Structure-Spectrum Relationships with Physically Organized Latent Spaces

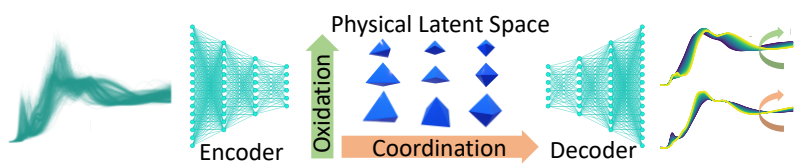

# Supporting Information for "Decoding Structure-Spectrum Relationships with Physically Organized Latent Spaces"

*Zhu Liang,*[1] *Matthew R. Carbone,*[2] *Wei Chen,*[1] *Fanchen Meng,*[1] *Eli Stavitski,*[3] *Deyu Lu,*[1,*] *Mark S. Hybertsen,*[1,*] *and Xiaohui Qu*[1*]

[1]Center for Functional Nanomaterials, Brookhaven National Laboratory, Upton, New York 11973, USA

[2]Computational Science Initiative, Brookhaven National Laboratory, Upton, New York 11973, USA

[3]National Synchrotron Light Source II, Brookhaven National Laboratory, Upton, New York 11973, USA

* dlu@bnl.gov, mhyberts@bnl.gov, xiaqu@bnl.gov

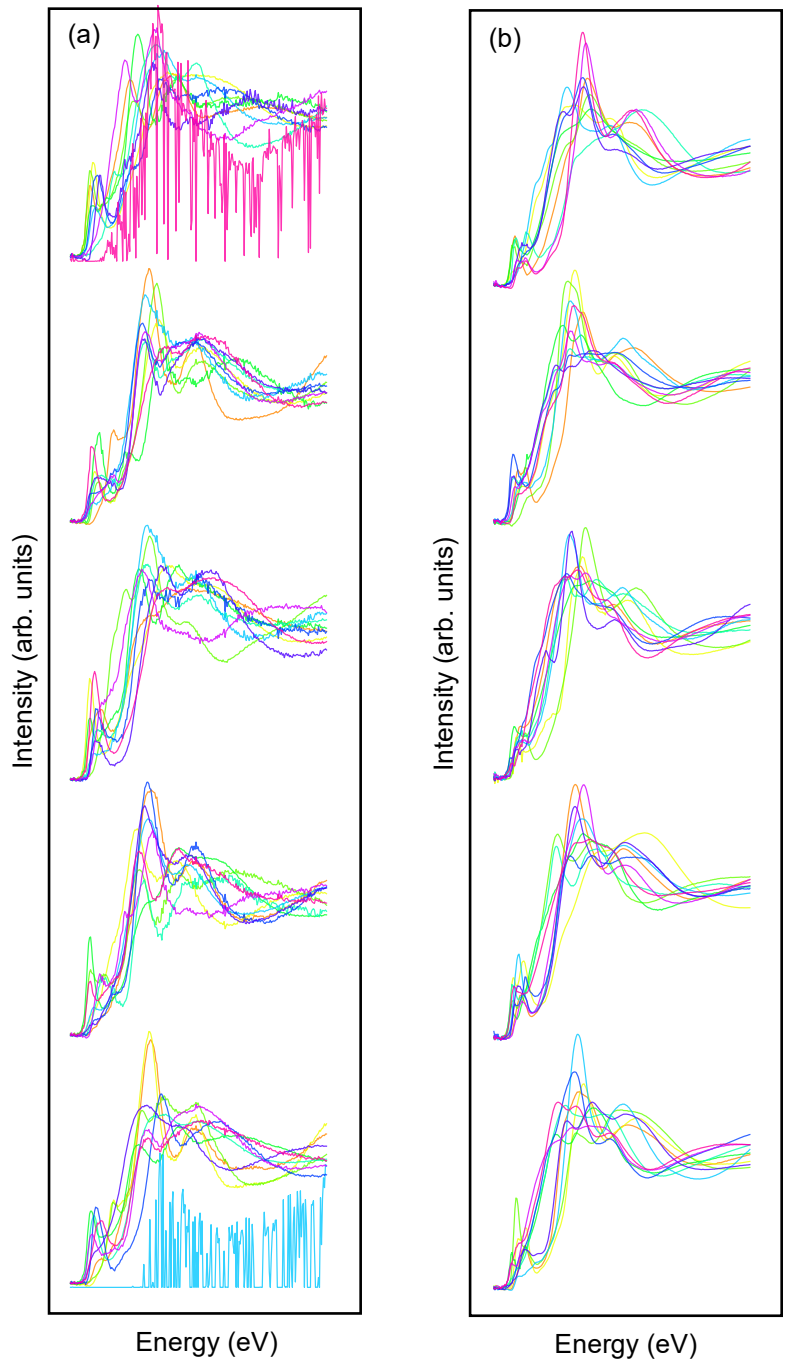

Figure S1. Comparing the generative characteristics for (a) a regular autoencoder (AE) model compared to (b) an adversarial autoencoder (AAE) model, each trained independently on the vanadium oxide data set. The dimension of the latent space is set to 6, representative of other models considered in this study. The decoded spectra are sampled in five cohorts of 10 randomly sampled data points in the latent space with each cohort displayed in a separate row. The main text highlights differences observed for AE versus AAE trained models in the literature. Here the consequence shortcomings of the AE model are illustrated for the exemplary XANES data specifically.

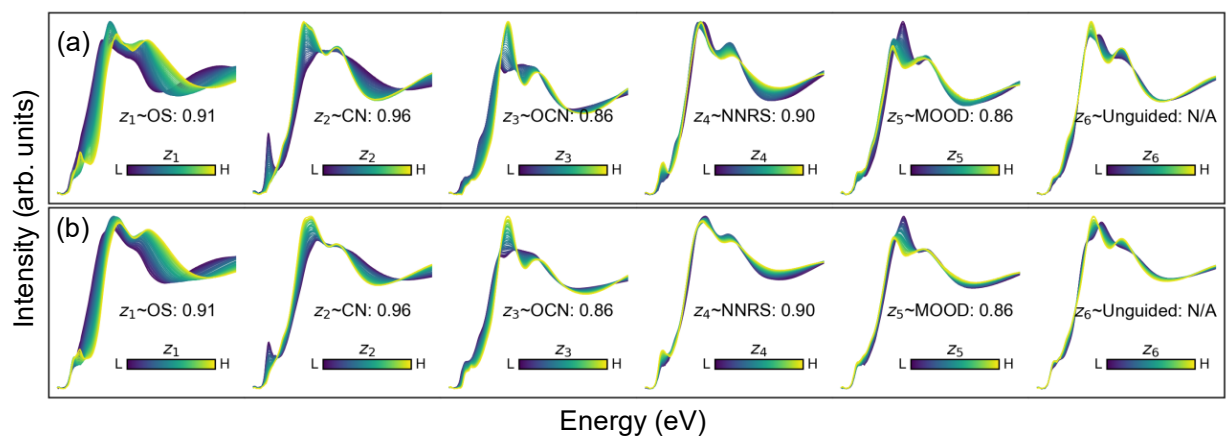

Figure S2. Spectral trends in vanadium oxides computed using the decoder in the RankAAE model by two different procedures: (a) Hold other latent dimensions to zero. (b) Sample and average other latent dimensions as described in Methods. Part (b) is duplicated from main text, Figure 5c.

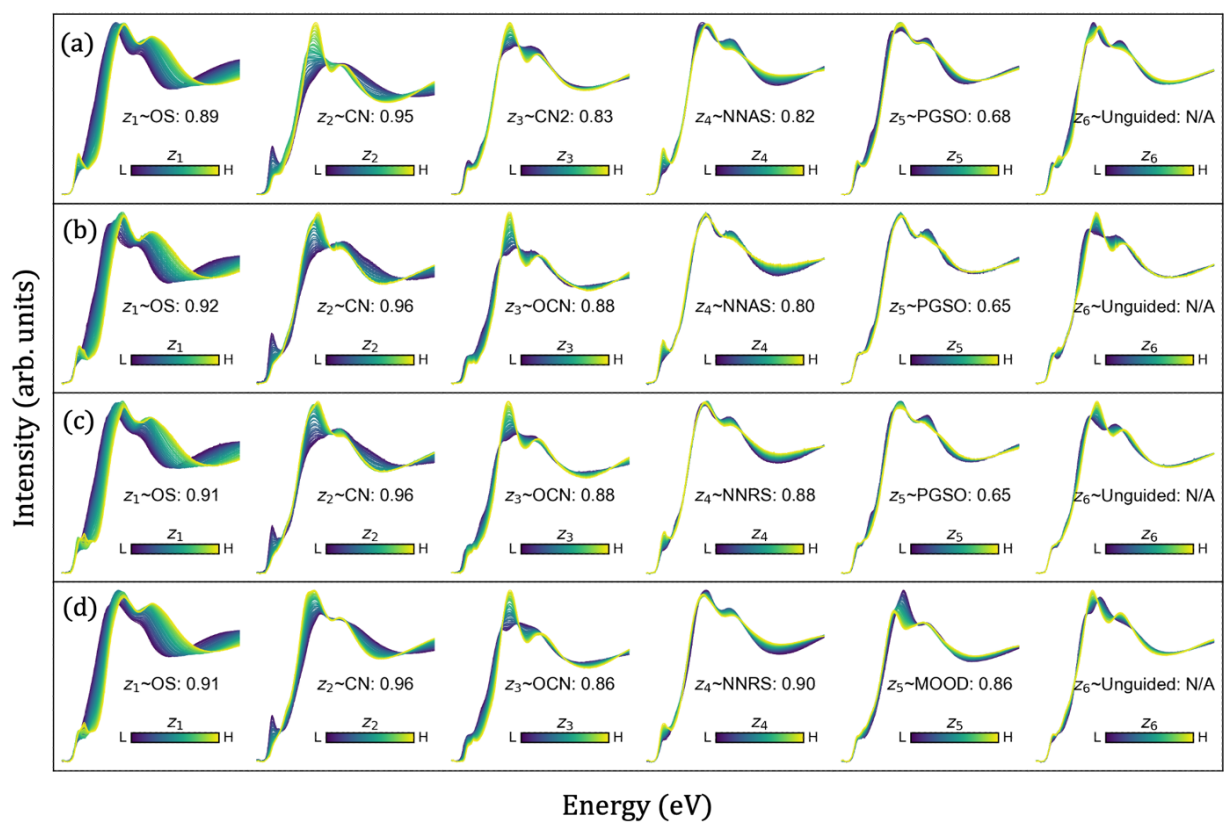

Figure S3. Spectral trends from a series of independently trained RankAAE models constrained with different sets of structure descriptors in each row as noted for each trend. Starting in the first row (a) with descriptors {CS, CN, CN-2, NNAS, PGSO}, progressively one descriptor is replaced to arrive at the final set {CS, CN, OCN, NNRS, MOOD} in the last row (d). The annotated label for each spectral trend refers to the related latent variable, structure descriptor, and the F1 score or the Spearman rank correlation coefficient (SRCC) between them.

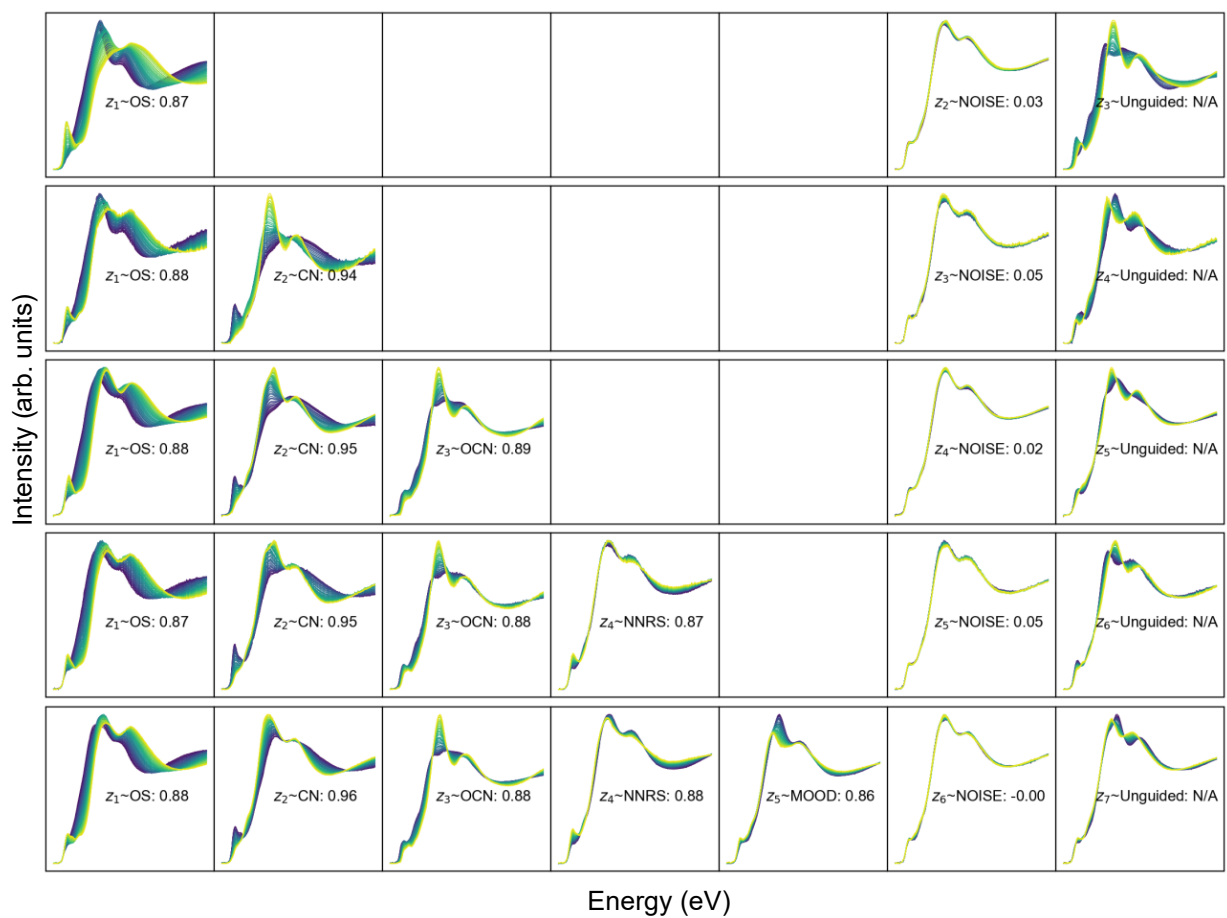

Figure S4. Series of independently trained RankAAE models showing the evolution of the models as structure descriptors are added sequentially, starting with OS. In addition, to the unguided latent space variable, we include another dimension guided to reproduce a pseudo-random number sequence as a structure descriptor (annotated as NOISE). In the top row, the dimension of the latent space is $N=3$. A dimension is added successively for each additional structure descriptor until $N=7$ in the final row.

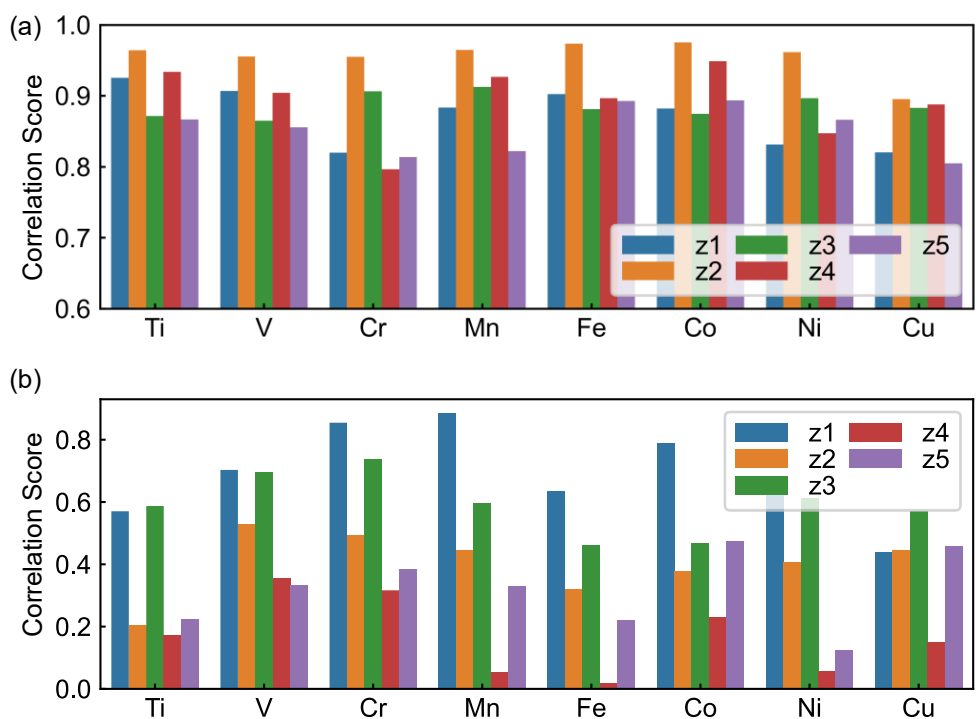

Figure S5. F1 score ($z_1$~OS and $z_2$~CN) and SRCC ($z_3$~OCN, $z_4$~NNRS and $z_5$~MOOD) assessing the correlation of latent variable to (a) the target structure descriptors and (b) the primary spectral descriptors for each of the RankAAE models trained independently for each transition metal oxide data set. The test portion of the data set is used for this assessment. The correlation between latent variable $z_5$ and spectral descriptor $E_{main}$ is computed for each CN category and the average value is shown in (b).

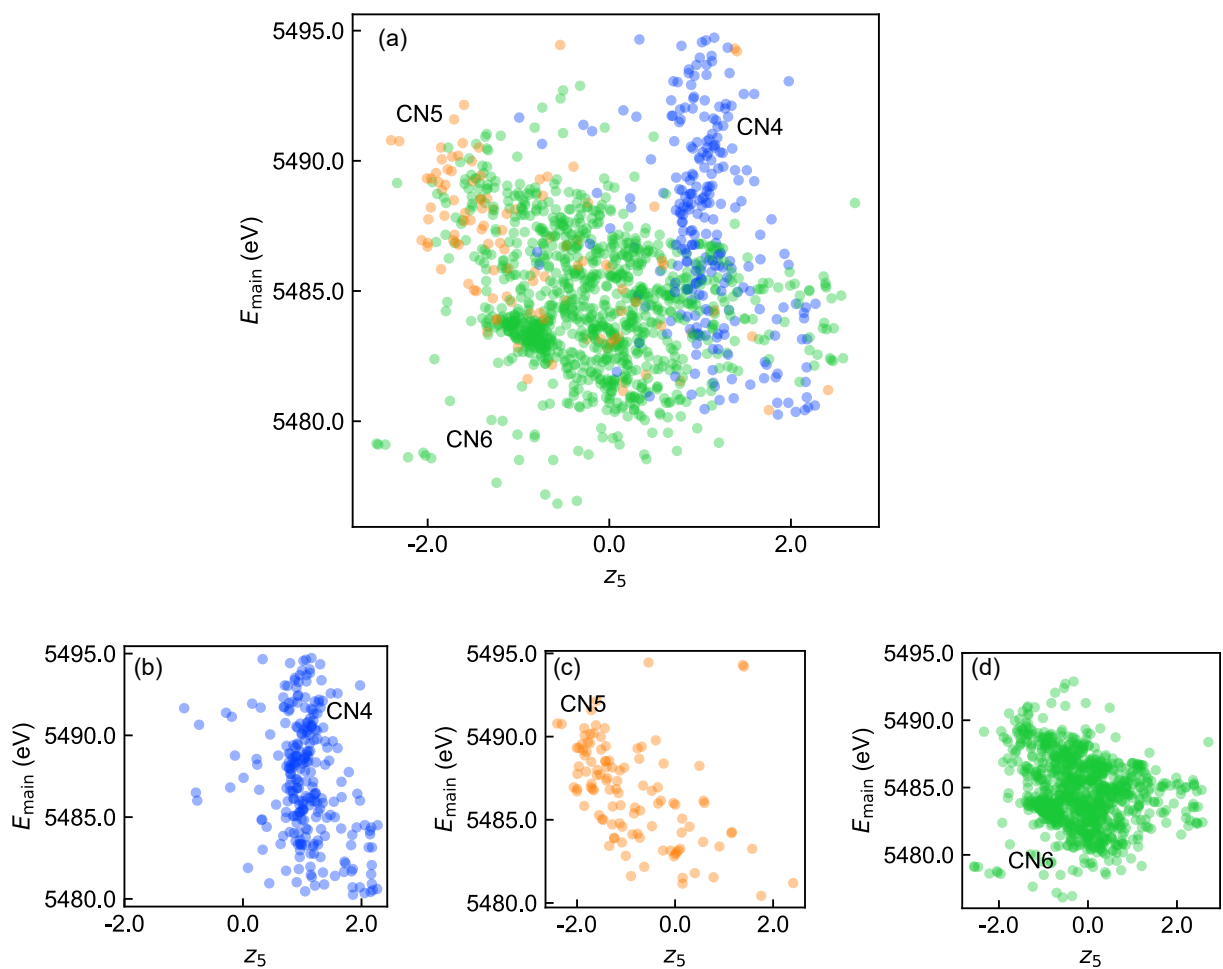

Figure S6. Illustration of multiple contributions to the main peak position spectral descriptor $E_{main}$ (y-axis) for the test set portion of the vanadium oxide data set. Contribution from CN is shown by color (CN4: blue, CN5: orange, CN6: green), and contribution from latent variable $z_5$ is shown by the x-axis. The top figure (a) shows the data in aggregate while the bottom role of figures (b-c) separates out the data according to CN.

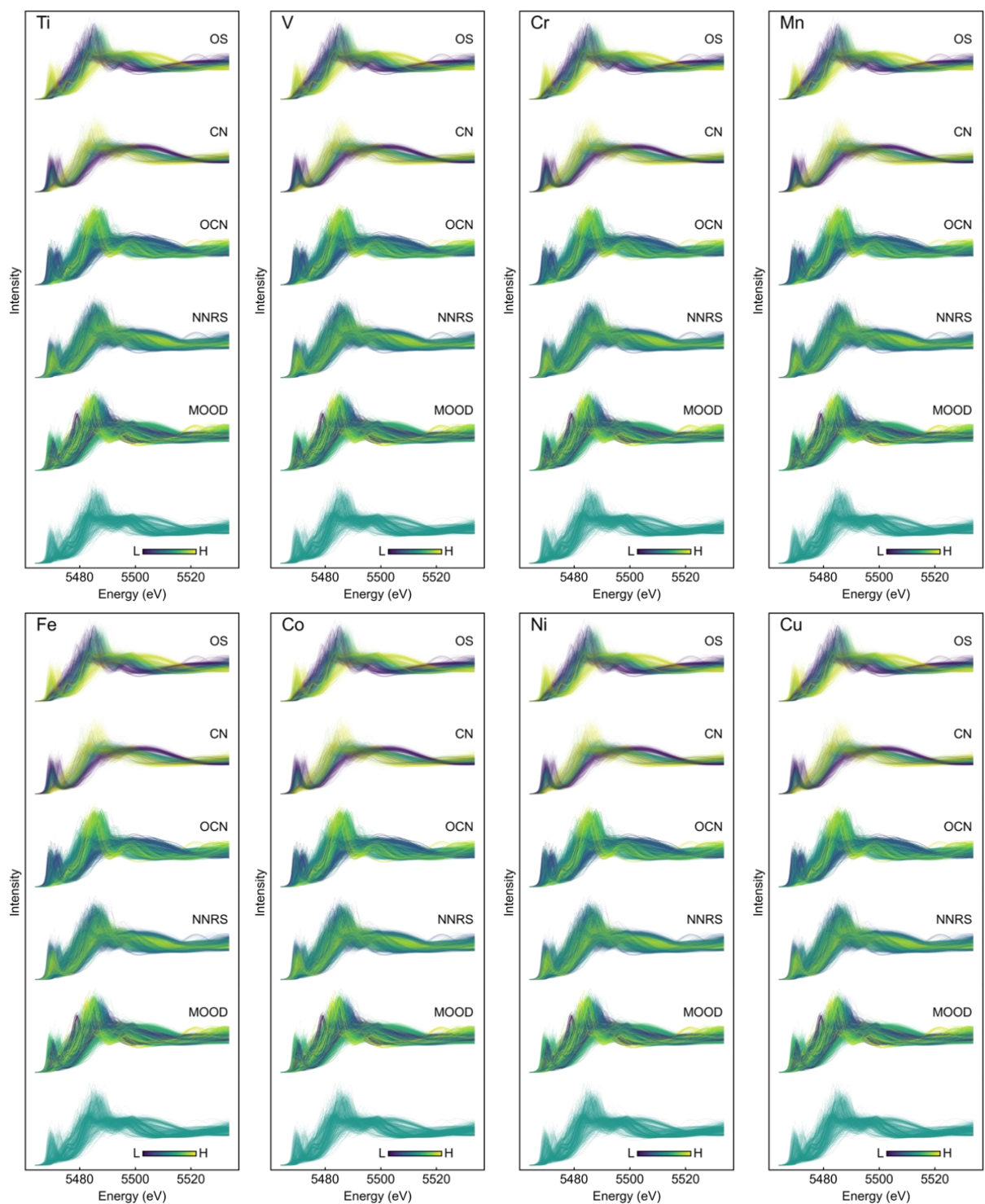

Figure S7. Full data set of K-edge XANES spectra for each of the transition metal oxides showing the full range of spectral variation. Each row is colored by a specific structure descriptor annotated in the plots: CS, CN, OCN, NNRS, MOOD. The bottom row is the full data set with no further distinction.
35

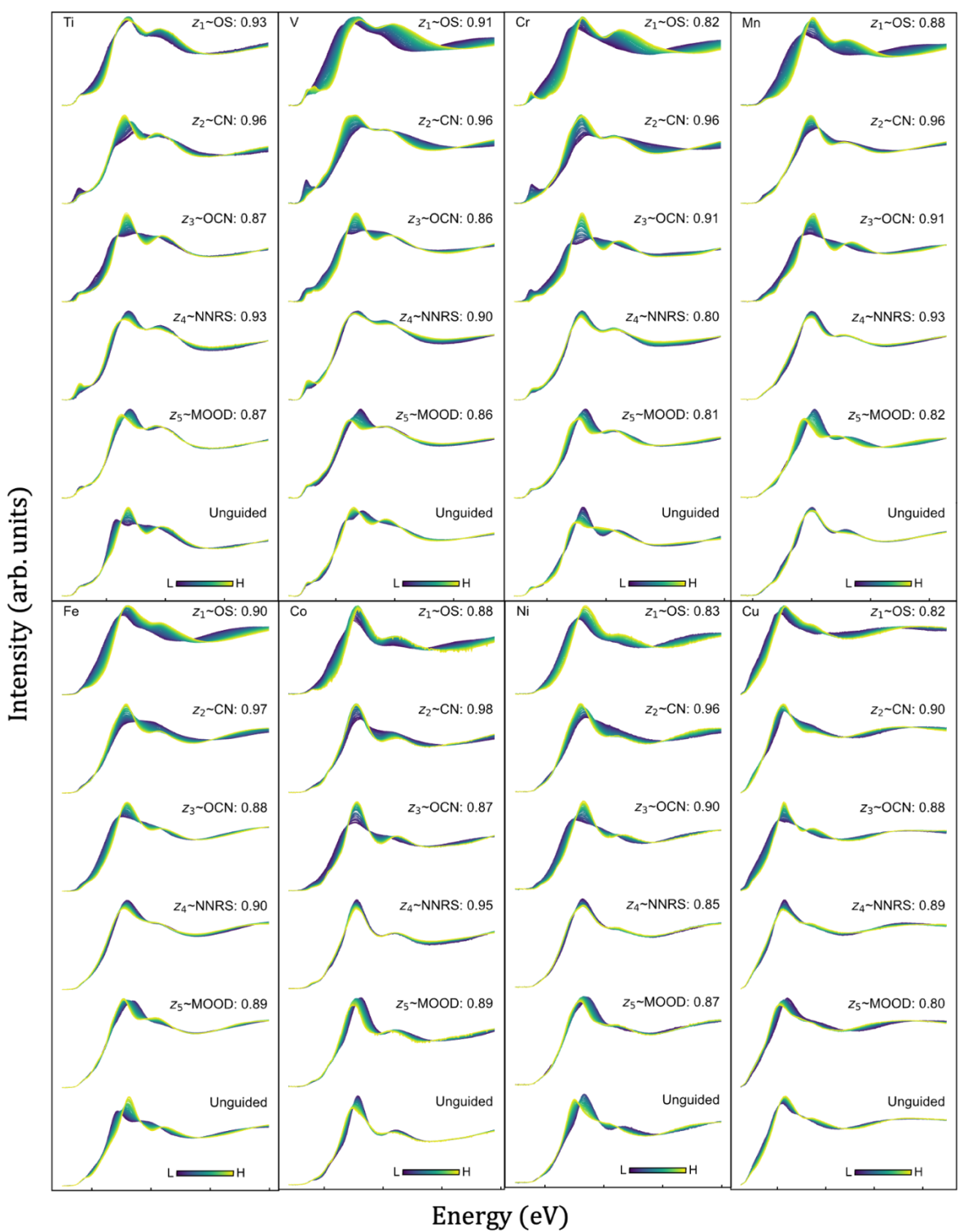

Figure S8. Spectral trends for independently trained RankAAE models for each of the transition metal oxide data sets using the final set of descriptors determined for the specific case of vanadium oxides and presented in Figure 5c. The format of the plot follows Figure 5c in the main text.

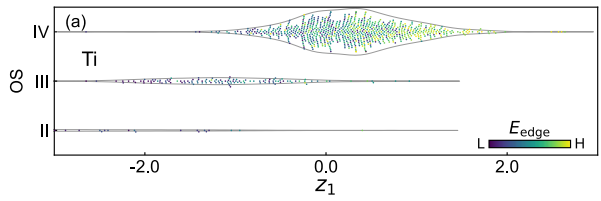
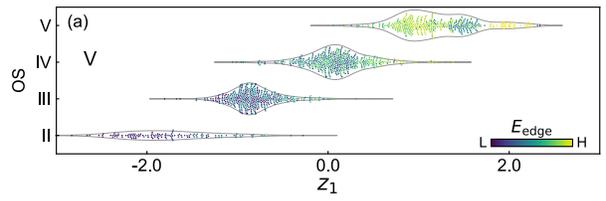
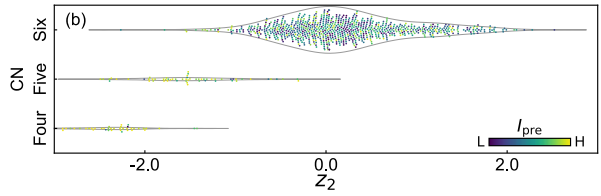
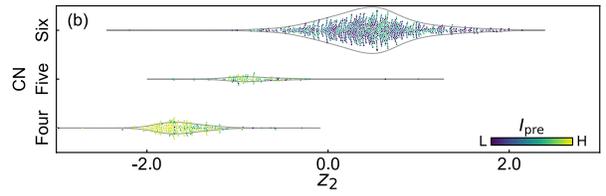
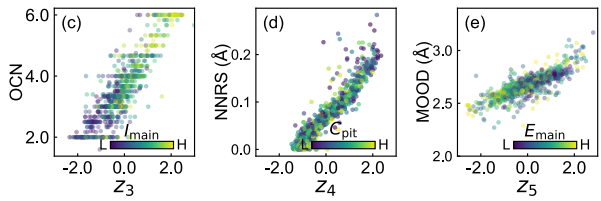
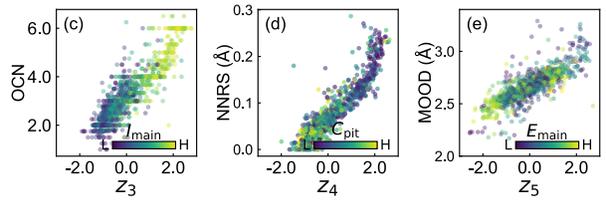
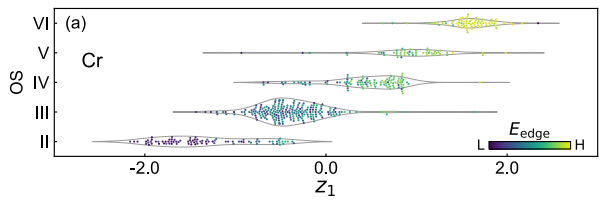
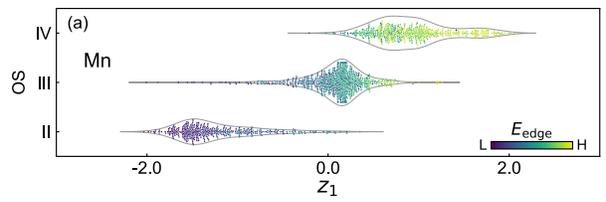
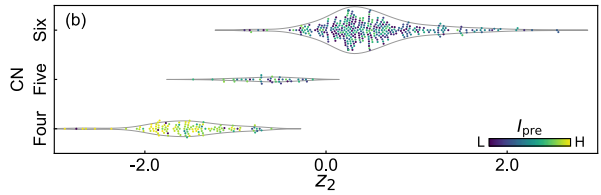
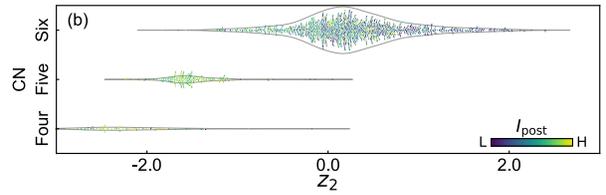
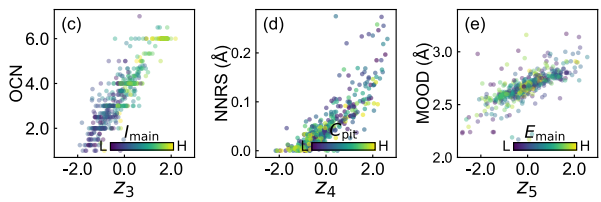
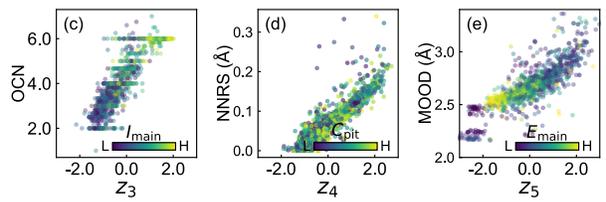

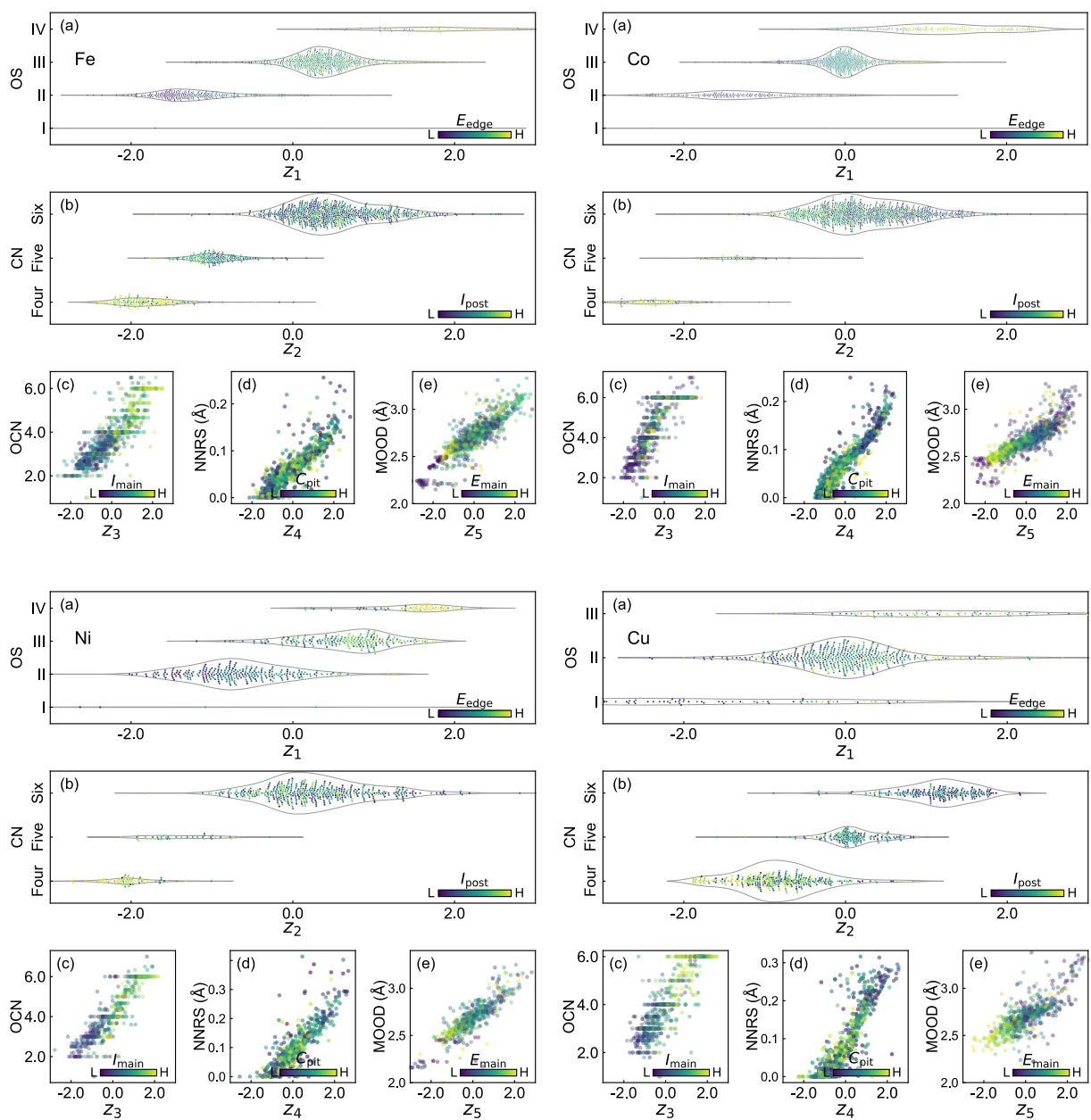

Figure S9. Correlation between the RankAAE latent space (x-axis), structure descriptor (y-axis), and spectral descriptor (color) for the each of the transition metal oxides data sets. The format of the plots follows Figure 6 in the main text.

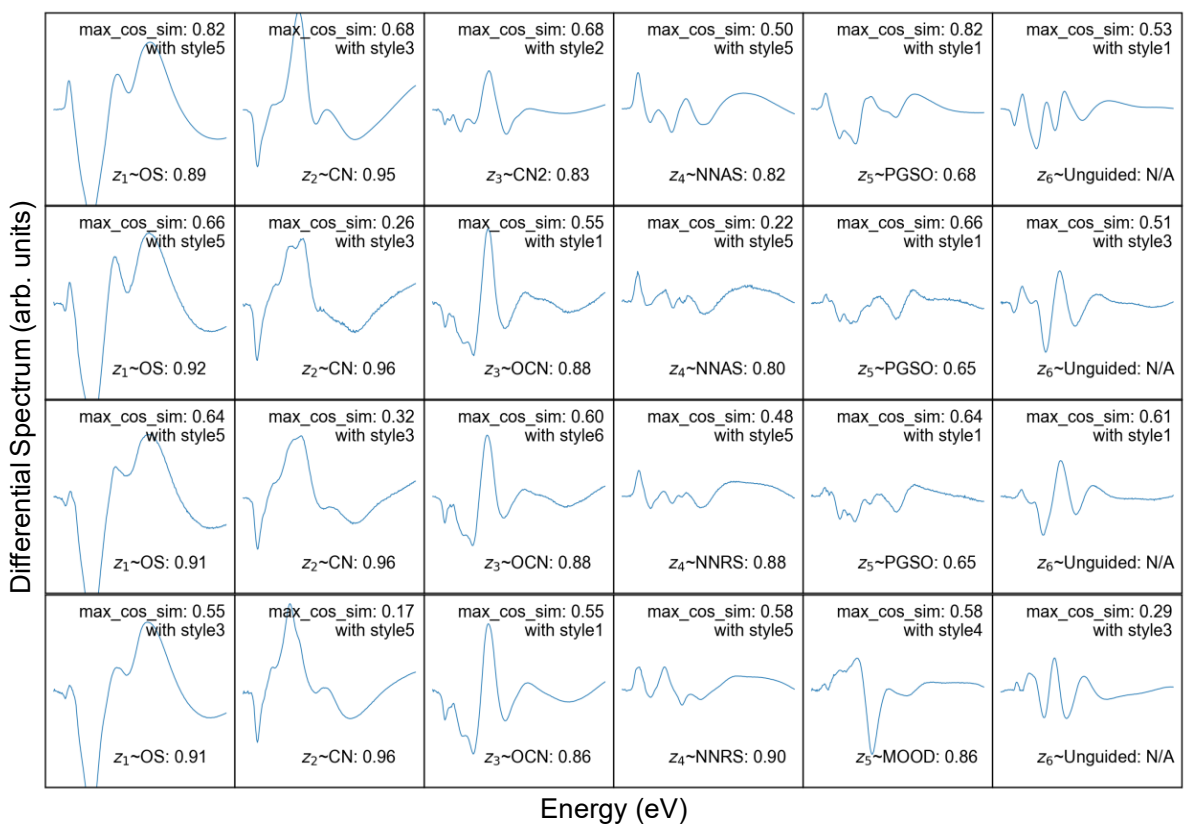

Figure S10. Alternative representation of the spectral fingerprint associated with each spectral trend derived from the sequence of RankAAE models constrained to different descriptors and shown in Figure S1. Shown is the difference between extremal spectra for each trend (95$^{th}$ and 5$^{th}$ percentile of the latent values). The scale for each box is identical, both amplitude (y-axis) and energy scale (x-axis), so the visual comparison of amplitude is meaningful. To assess overlaps, a matrix of cosine similarities is computed for the fingerprints in each row. The value of the largest overlap is annotated by the "max_cos_sim" score and the specific style with the largest overlap is noted.

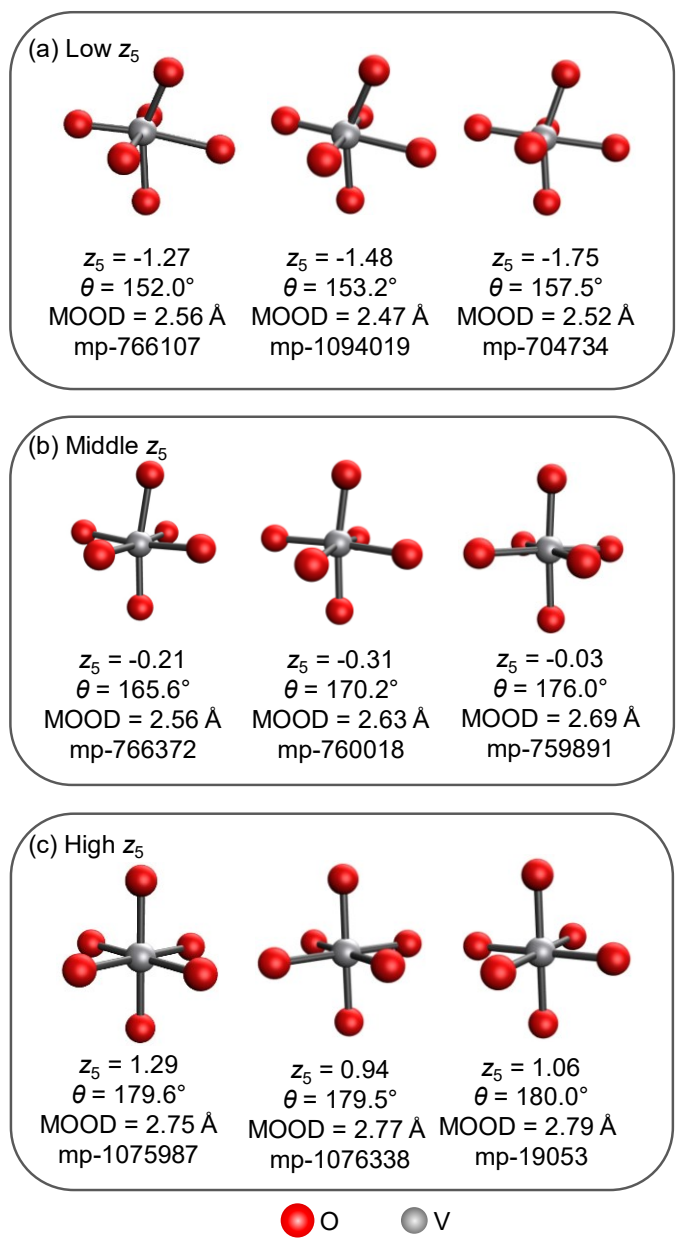

Figure S11. Example atomic motifs associated with different ranges of latent variable $z_5$ value. The material project ID, value of latent variable $z_5$, angle between the cation bonds to the two axial oxygen ($\theta$), and structure descriptor MOOD are annotated below each motif.

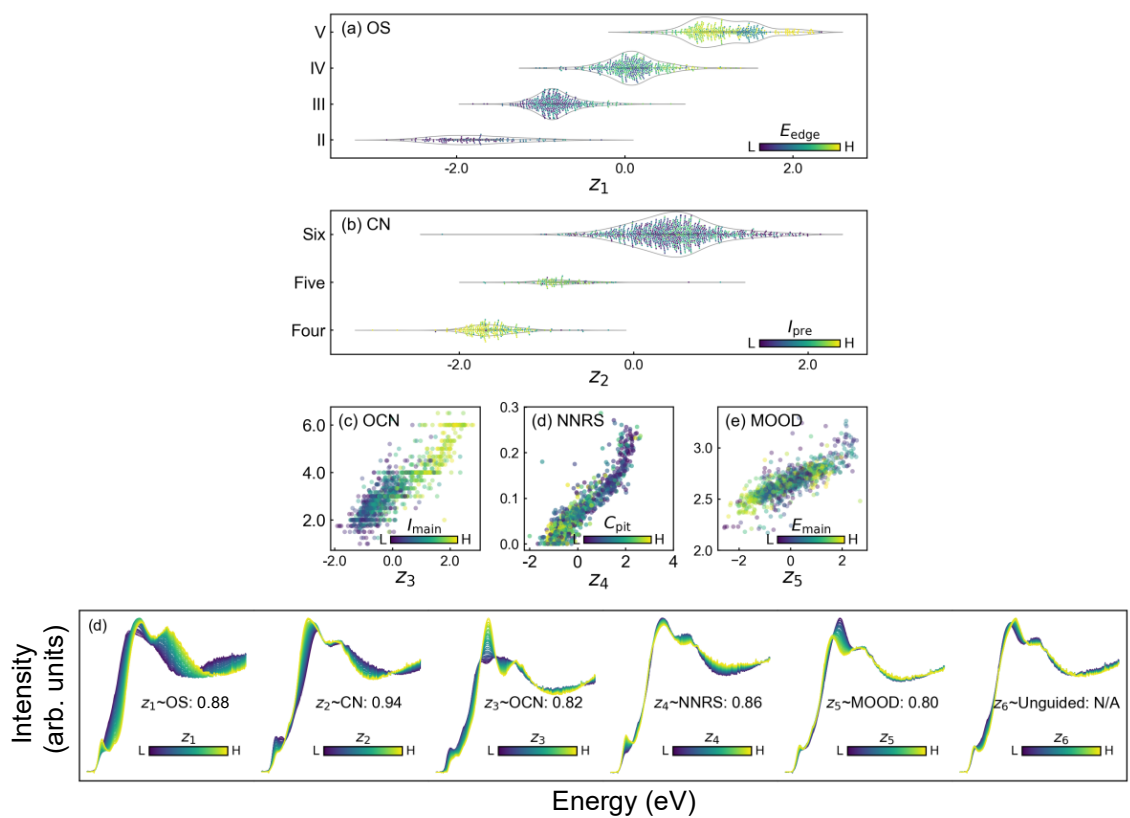

Figure S12. For each choice of descriptor set, the RankAAE model is trained 100 times. The performance score $S$ defined by Methods equation (2) on the test portion of the data set ranges from -12.91 to 7.02. The correlation plots are shown for the worst model with $S$=-12.91 for vanadium oxide.

Table S1. Size of datasets for all metals.

|    | Total | Training | Test | Validation |
|----|-------|----------|------|------------|
| Ti | 6488  | 4541     | 973  | 974        |
| V  | 9312  | 6518     | 1396 | 1398       |
| Cr | 3596  | 2517     | 539  | 540        |
| Mn | 11755 | 8228     | 1763 | 1764       |
| Fe | 7446  | 5212     | 1116 | 1118       |
| Co | 10146 | 7102     | 1521 | 1523       |
| Ni | 3666  | 2566     | 549  | 551        |
| Cu | 4564  | 3194     | 684  | 686        |

Table S2. Statistical characterization of performance metrics across 100 independently trained models for the case of vanadium oxide. The model selection process utilizes values further normalized by z-scores (Methods). Metrics are computed on the test portion of the data set.

|  | Average | Standard Deviation | Minimum | Maximum |
|---|---|---|---|---|
| $\rho_{ij}$ | 0.277 | 0.092 | 0.049 | 0.556 |
| Reconstruction Error* | 0.037 | 0.001 | 0.034 | 0.041 |
| $\rho'_1: z_1\sim$OS | 0.890 | 0.008 | 0.876 | 0.907 |
| $\rho'_2: z_2\sim$CN | 0.946 | 0.010 | 0.911 | 0.964 |
| $\rho'_3: z_3\sim$OCN | 0.860 | 0.011 | 0.821 | 0.881 |
| $\rho'_4: z_4\sim$NNRS | 0.884 | 0.016 | 0.842 | 0.910 |
| $\rho'_5: z_5\sim$MOOD | 0.840 | 0.012 | 0.805 | 0.863 |

* The reconstruction error is computed as the mean absolute deviation (MAD) between the original spectrum and reconstructed spectrum